\documentclass[aps,superscriptaddress,asmsymb,twocolumn,showpacs,amsmath,prl,showkeys]{revtex4-1}

\usepackage{graphicx}
\usepackage{subfigure}
\usepackage{epsfig}
\usepackage{dcolumn}
\usepackage{bm}
\usepackage{color}
\usepackage{xcolor}
\usepackage{bm}
\usepackage{hyperref}
\usepackage{amsmath, amssymb, dsfont}

\def\abs#1{\left|#1\right|}

\def\sgn{{\rm sgn}}

\def\be{\begin{equation}}       \def\ee{\end{equation}}
\def\bea{\begin{eqnarray}}      \def\eea{\end{eqnarray}}
\def\ba{\begin{array}}
\def\ea{\end{array}}
\def\bnum{\begin{enumerate} }
\def\enum{\end{enumerate}}

\def\=>{\Rightarrow}
\def\>{\rightarrow}

\def\eye2{Fathbb{I}}
\def\bk{{\bf k}}
\def\bv{{\bf v}}

\def\br{{\bf r}}

\renewcommand{\v}[1]{{\bf #1}}

\renewcommand{\>}{\rangle}

\newcommand{\e}{\epsilon}

\begin{document}

\title{A semiclassical approach to surface Fermi arcs in Weyl semimetals}
\author{Jiajia Huang}
\affiliation{State Key Laboratory of Optoelectronic Materials and Technologies, Guangdong Provincial Key Laboratory of Magnetoelectric Physics and Devices, School of Physics, Sun Yat-sen University, Guangzhou 510275, China}
\author{Luyang Wang}
\email{wangly@szu.edu.cn}
\affiliation{College of Physics and Optoelectronic Engineering, Shenzhen University, Shenzhen 518060, China}
\author{Dao-Xin Yao}
\email{yaodaox@mail.sysu.edu.cn}
\affiliation{State Key Laboratory of Optoelectronic Materials and Technologies, Guangdong Provincial Key Laboratory of Magnetoelectric Physics and Devices, School of Physics, Sun Yat-sen University, Guangzhou 510275, China}
\affiliation{International Quantum Academy, Shenzhen 518048, China}

\begin{abstract}
We present a semiclassical explanation for the morphology of the surface Fermi arcs of Weyl semimetals. Viewing the surface states as a two-dimensional Fermi gas subject to band bending and Berry curvatures, we show that it is the non-parallelism between the velocity and the momentum that gives rise to the spiral structure of Fermi arcs. We map out the Fermi arcs from the velocity field for a single Weyl point and a lattice with two Weyl points. We also investigate the surface magnetoplasma of Dirac semimetals in a magnetic field, and find that the drift motion, the chiral magnetic effect and the Imbert-Fedorov shift are all involved in the formation of surface Fermi arcs. Our work not only provides an insightful perspective on the surface Fermi arcs and a practical way to find the surface dispersion, but also paves the way for the study of other physical properties of the surface states of topological semimetals, such as transport properties and orbital magnetization, using semiclassical methods.
\end{abstract}
\date{\today}

\keywords{Weyl semimetals, Surface Fermi arcs, Semiclassical equations of motion, Dirac semimetals, Surface magnetoplasma}

\pacs{71.20.Gj, 73.20.-r, 73.20.Mf, 03.65.Vf}

\maketitle

{\it Introduction.}---Weyl semimetals (WSMs) have distinct surface states that form Fermi arcs in the surface Brillouin zone. The existence of surface Fermi arcs is a characteristic property of WSMs, which can be understood in several ways. Firstly, in each slice between a pair of Weyl points which carry opposite monopole charges, quantum anomalous Hall effect (QAHE) occurs and the slice can be viewed as a Chern insulator, which possesses a gapless edge state crossing the Fermi energy at a single point of the surface Brillouin zone. Connecting all the points yields a Fermi arc, which ends at the projections of the bulk Weyl points\cite{Wan2011}. An alternative way to understand the Fermi arcs is from the quantum mechanical solution of Weyl equation. The boundary condition of Weyl equation tells that at the surface, the surface states must break the rotational symmetry of Weyl equation, resulting in Fermi arc surface states that exponentially decay as they go into the bulk\cite{Witten2016}. Another viewpoint is to start from a two-dimensional system with a closed Fermi pocket and grow it into a three-dimensional bulk\cite{Devizorova2017}. The Fermi pocket splits into two halves, migrating to opposite surfaces as the separation between the surfaces increases. Other studies of Fermi arcs also exist\cite{Burkov2011,Ojanen2013,Haldane2014,Okugawa2014,Sun2015, Kim2016}. Surface Fermi arcs have been observed in WSMs\cite{Xu2015Science,Lv2015PRX,Xu2015NP,Huang2015,Yang2015,Deng2016NP}, Weyl metamaterials\cite{Lu2015,Cheng2020} and can also exist in circuit lattices\cite{Weng2018Re}. Besides Fermi arcs, other characteristic phenomena of WSMs, such as quantum anomalous Hall effect\cite{Yang2011} and chiral anomaly\cite{Aji2012,Son2013,Xiong2015,Hosur2013}, have also been observed.

Here, we adopt a semiclassical approach to interpret the morphology of surface Fermi arcs of WSMs in the presence of band bending. We first prove that for an isotropic Weyl point, the surface states cannot have a closed Fermi pocket due to the non-parallelism between the total velocity, which is the sum of the band velocity and anomalous velocity, and the momentum of any surface state. Then we extend the proof to anisotropic and (or) tilted Weyl points, and argue that the Fermi surface has a spiraling structure. The spiraling Fermi arcs are computed explicitly by semiclassical equations of motion (EOMs) for a single Weyl point and a lattice system with two Weyl points. We also investigate the surface states of Dirac semimetals (DSMs) in a magnetic field, which are called the surface magnetoplasma. In analogy with the quantum Hall effect (QHE), the magnetoplasma is confined to the surface by the combined effect of a confining potential and the magnetic field, and has chiral nature from two origins, the drift motion and the chiral magnetic effect.

The semiclassical approach has also been used to study the Imbert-Fedorov (IF) shift of Weyl fermions\cite{Jiang2015,Yang2015PRL,Wang2017}: when a wavepacket of Weyl fermions with a certain chirality is incident on an interface or surface and reflected, it gains a transverse shift. While the IF shift of Weyl fermions is a bulk phenomena, we show that it is closely related to the surface Fermi arcs.

{\it Semiclassical approach to the formation of Fermi arcs.}---The semiclassical EOMs read\cite{Xiao2010}
\begin{eqnarray}
   \dot{\br}&=&\nabla_\bk \mathcal{E}(\bk)-\dot{\bk}\times{\bf\Omega}(\bk),\label{eq:velocity}\\
   \dot{\bk}&=&-\nabla_\br U(\br)-\dot{\br}\times{\bf B},\label{eq:force}
\end{eqnarray}
where we have set $\hbar=1$ and $e=1$. $\mathcal{E}(\bk)$ is the band dispersion, $U(\br)$ is the potential, ${\bf\Omega}(\bk)$ is the Berry curvature and $\bf B$ is an external magnetic field.

\begin{figure}[t]
  \centering
  \includegraphics[width=0.4\textwidth]{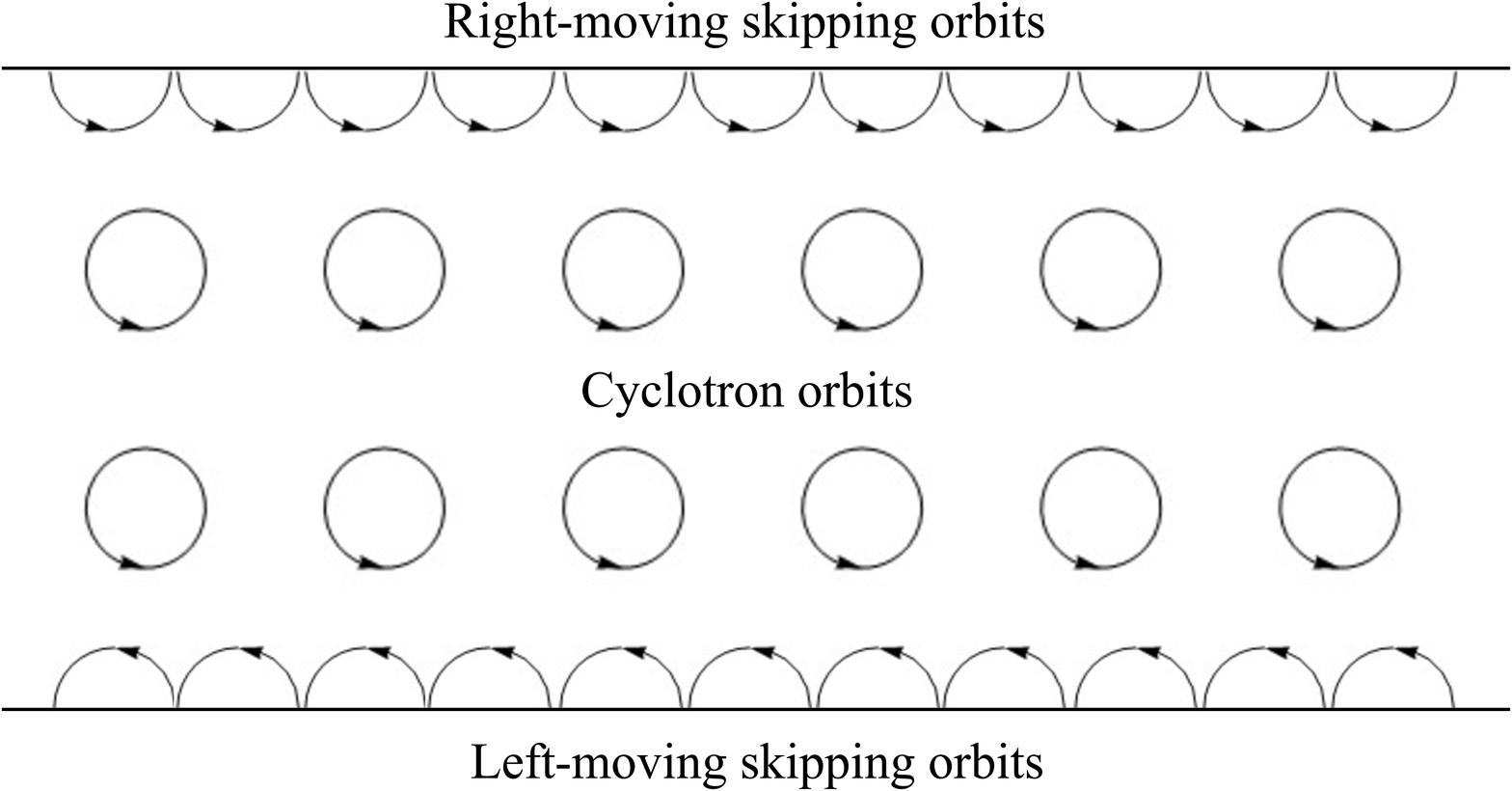}\\
  \caption{Semiclassical orbits in the quantum Hall effect.}\label{fig:QHE}
  \end{figure}

Let us review what we can get from the semiclassical EOMs for the QHE and the QAHE of two-dimensional systems. In the QHE of a free electron gas, an external magnetic field is applied, ${\bf B}\neq0$ and ${\bf\Omega}(\bk)=0$. In the bulk, the EOMs reduce to $\dot{\br}=\bk/m$ and $\dot{\bk}=-\dot{\br}\times{\bf B}$ with $m$ being the electron mass, which lead to cyclotron motion in both real space and momentum space. At the boundary of a sample, the steep confining potential $U(\br)$ results in a large electric field, reversing the momentum in the direction normal to the boundary, and only a portion of the cyclotron orbit is completed. Then another orbit starts with the reflected momentum as its initial momentum. All these ``skipping orbits" compose the chiral edge state of the QHE, as shown in Fig.\ref{fig:QHE}. The QAHE is slightly different\cite{Sinitsyn2007}. In this case, ${\bf B}=0$ while ${\bf\Omega}(\bk)$ is in the direction perpendicular to the system. The anomalous velocity, i.e. the second term in Eq.\ref{eq:velocity}, is perpendicular to the gradient of the confining potential, giving rise to the chiral edge state.
The helical edge states of quantum spin Hall effect have also been understood semiclassically\cite{Shi2013,Ferreira2018}. Naively, one may infer that in WSMs, since each slice between a pair of Weyl points can be viewed as a Chern insulator (QAHE), the semiclassical orbits make no difference with that in a real Chern insulator. However, since $\bf\Omega$ has three components in WSMs instead of one in a Chern insulator, other components can change the morphology of the surface Fermi arcs.

\begin{figure}[t]
  \centering
  \subfigure[]{\includegraphics[width=0.2\textwidth]{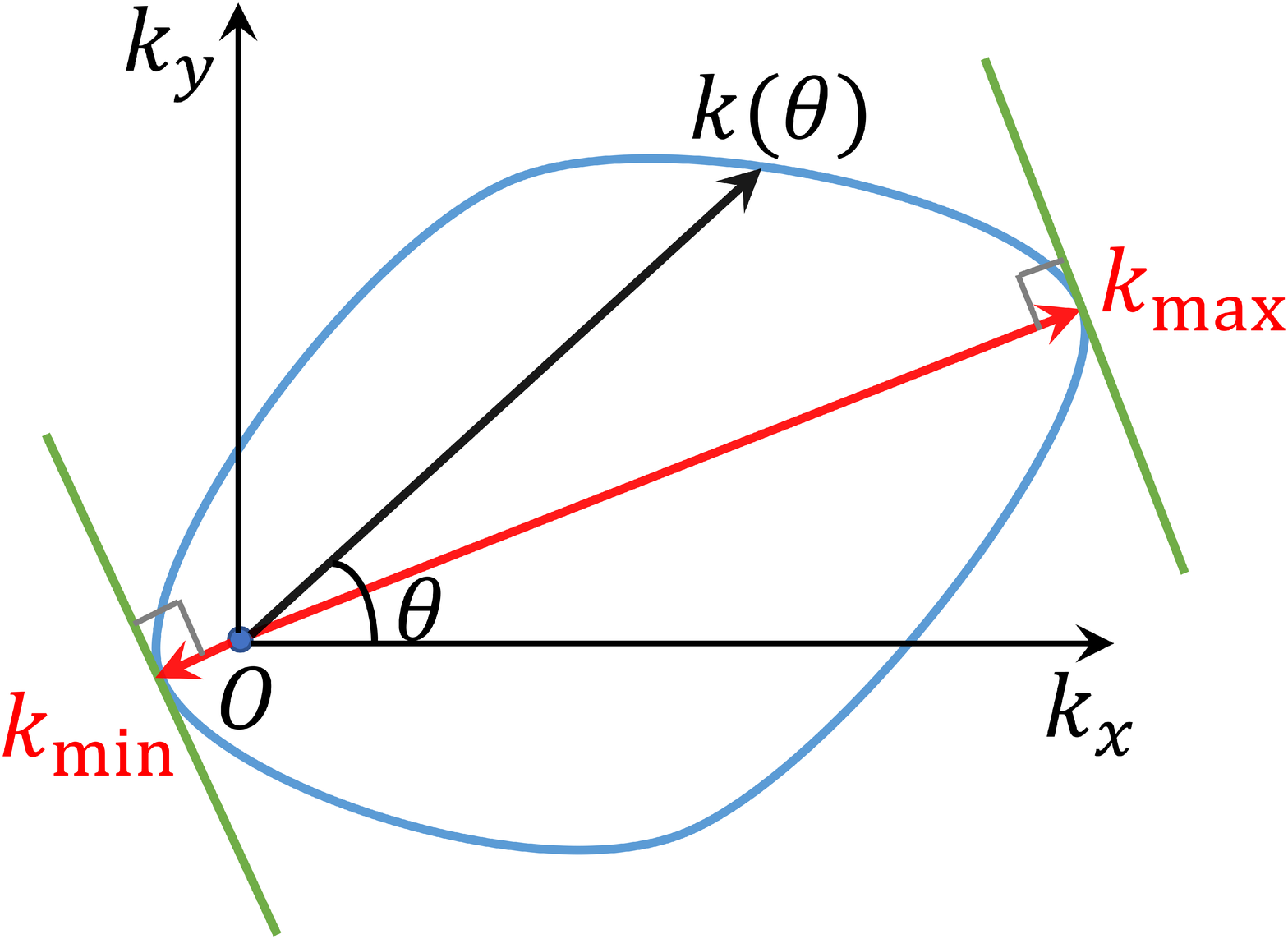}\label{fig:closed}}~~
  \subfigure[]{\includegraphics[width=0.2\textwidth]{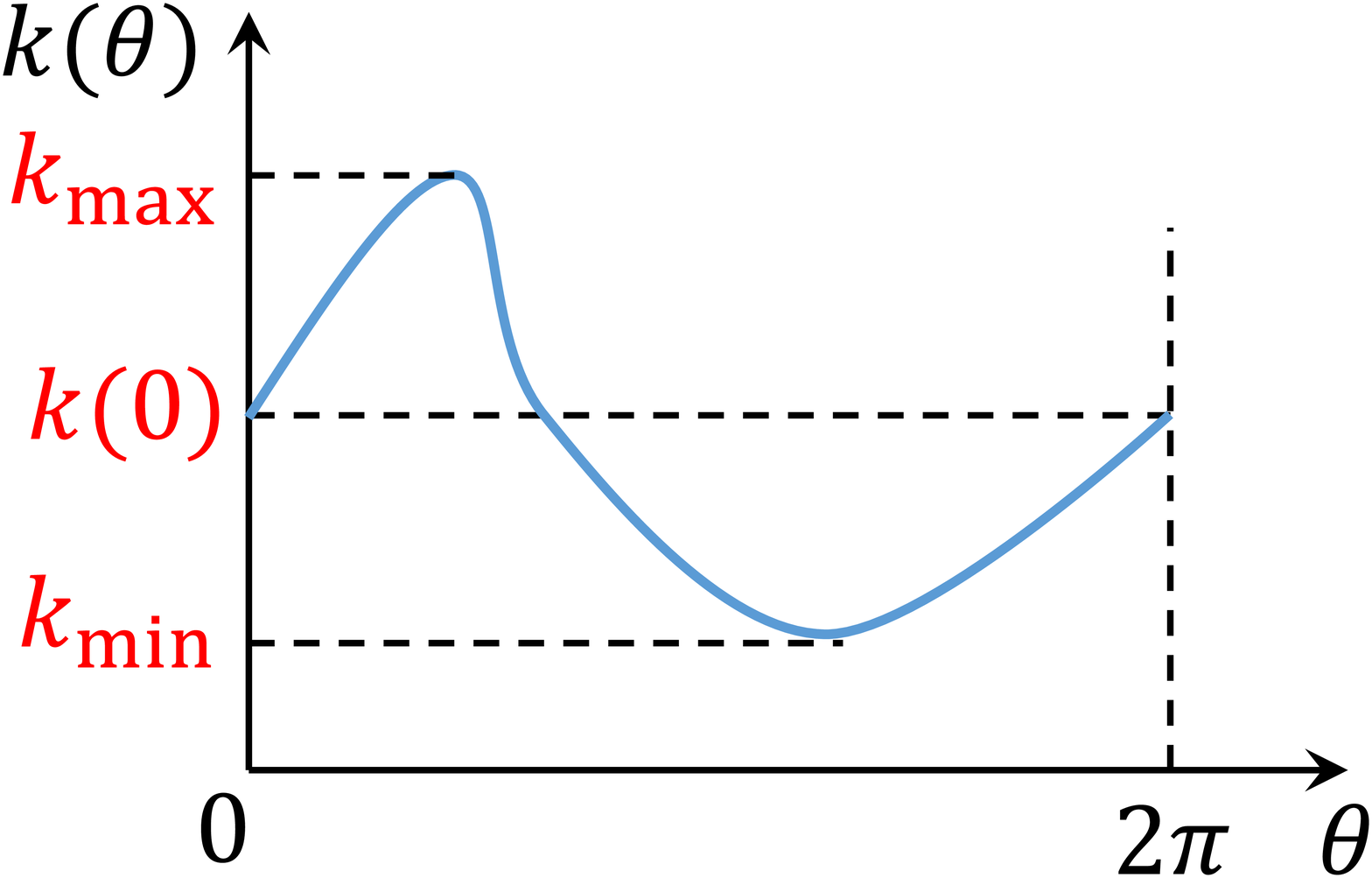}\label{fig:k_theta}}
  \subfigure[]{\includegraphics[width=0.2\textwidth]{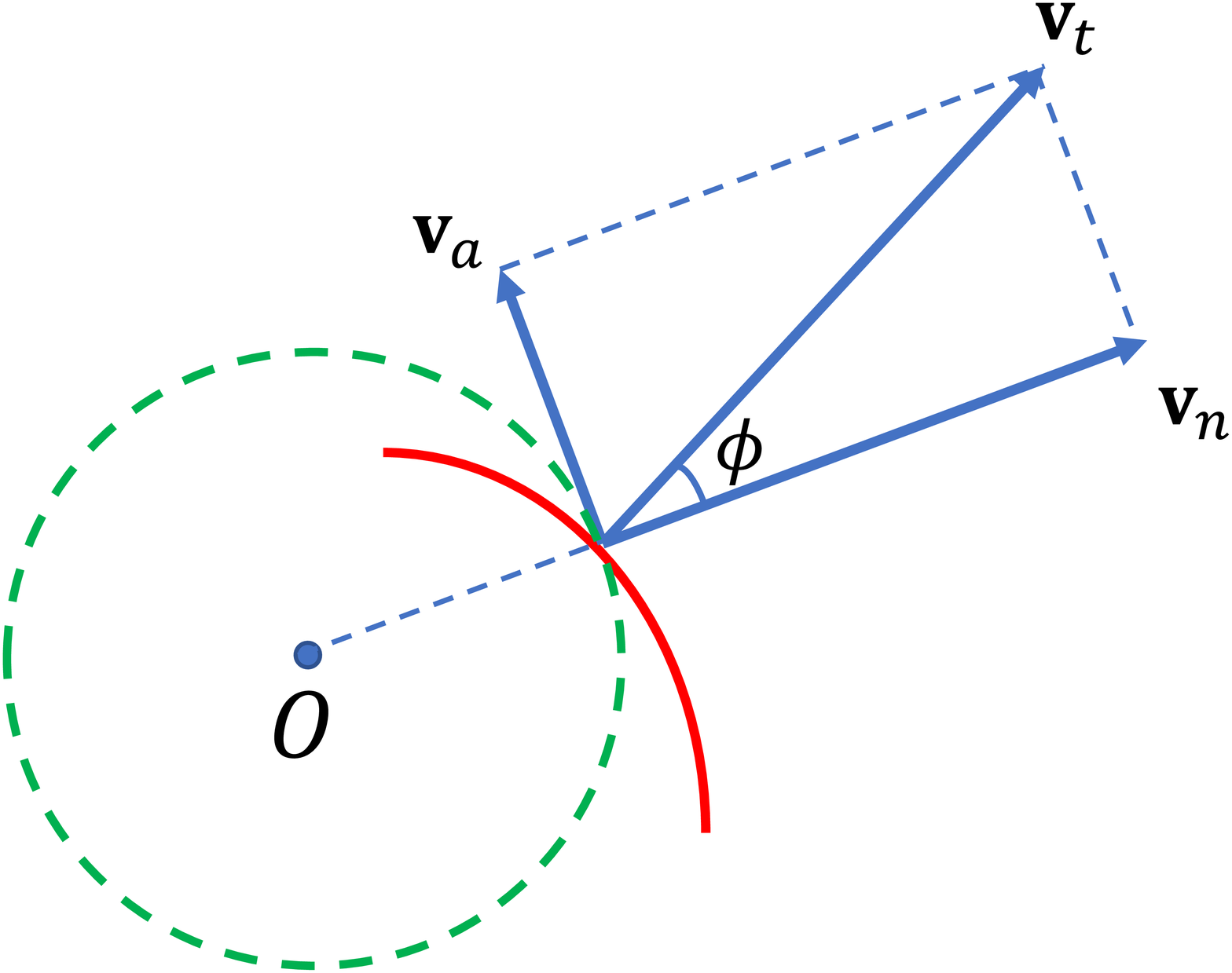}\label{fig:phi}}~~
  \subfigure[]{\includegraphics[width=0.2\textwidth]{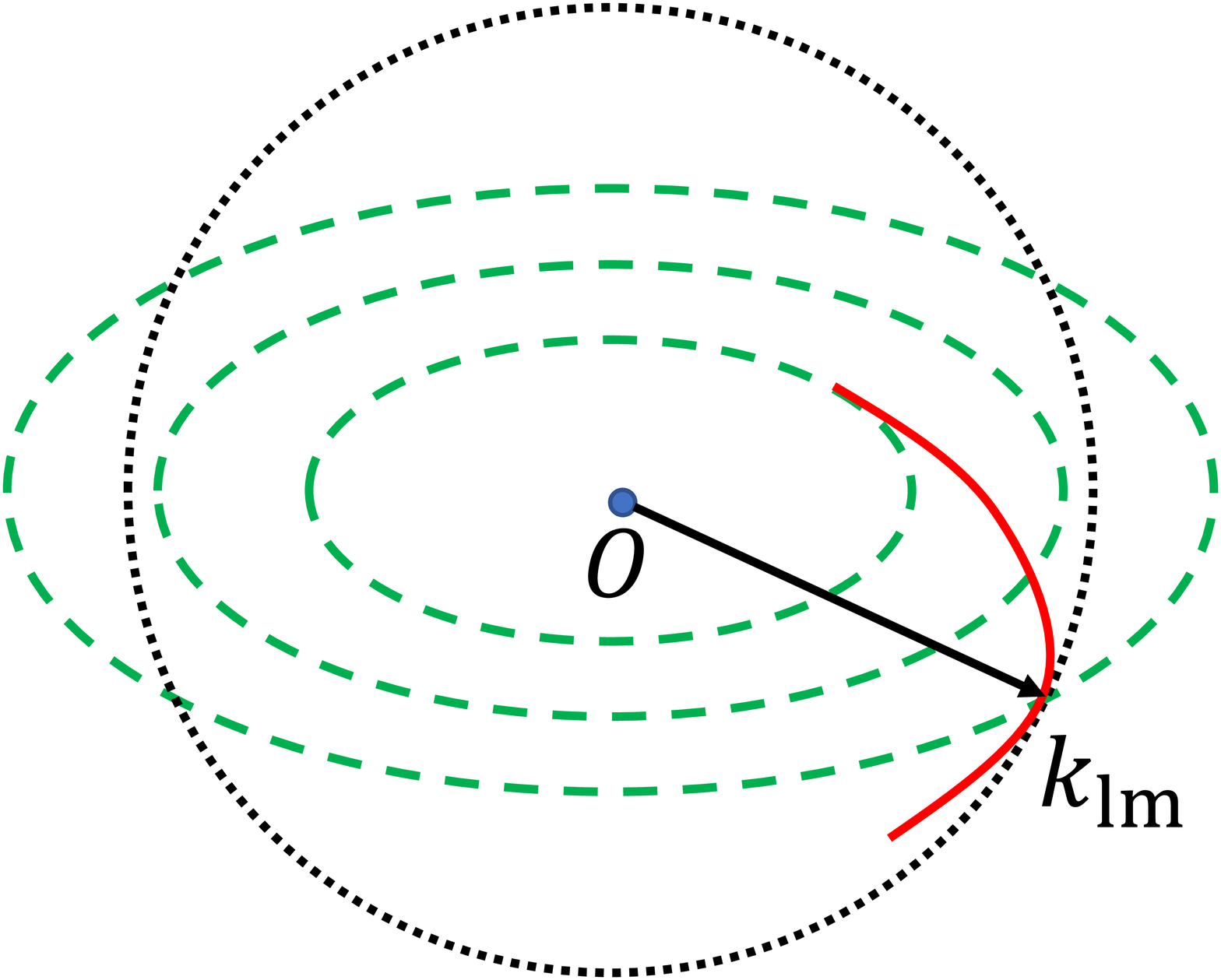}\label{fig:ellipse}}
  \subfigure[]{\includegraphics[width=0.2\textwidth]{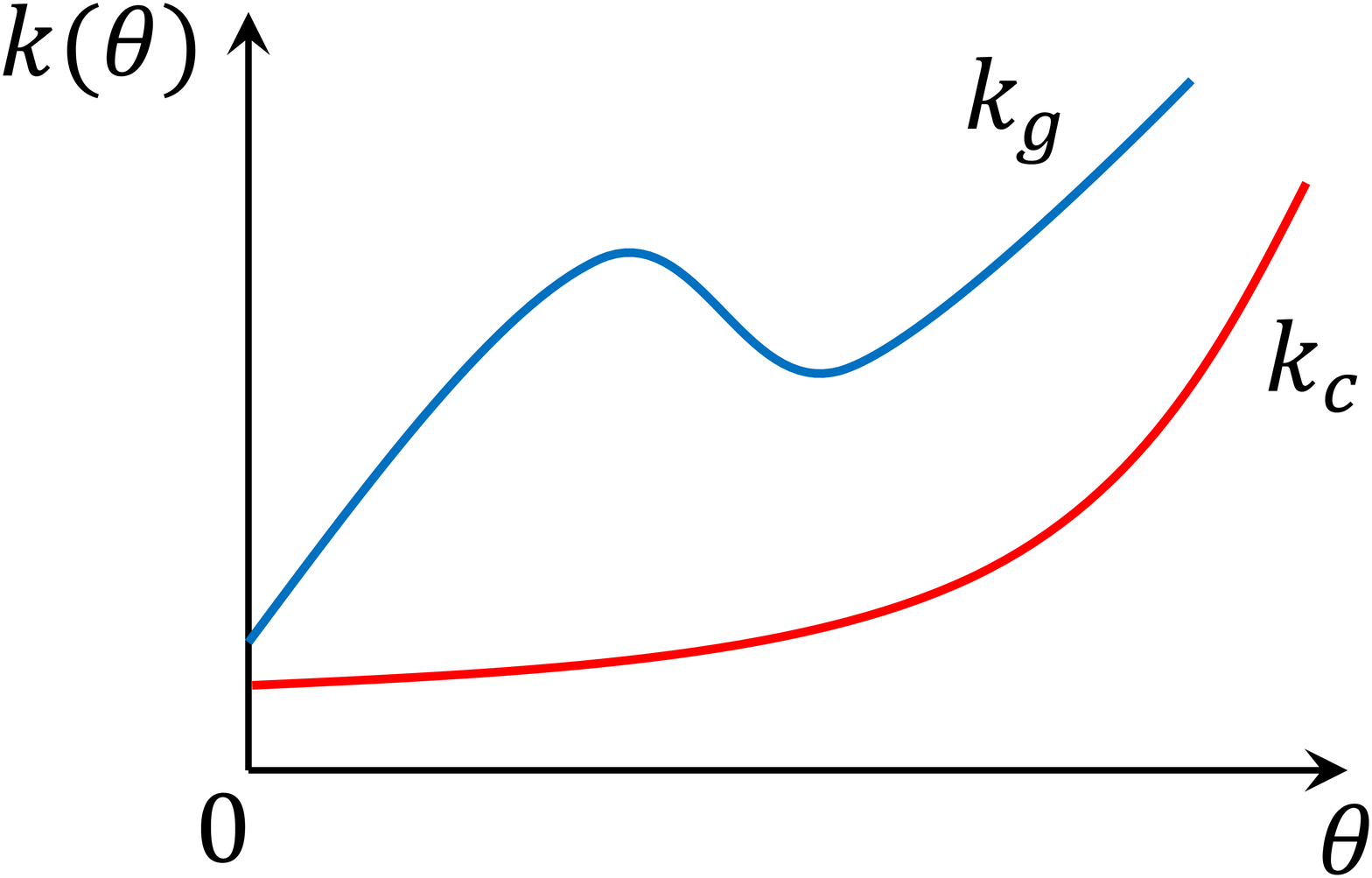}\label{fig:kg}}~~~~~~~~~
  \subfigure[]{\includegraphics[width=0.16\textwidth]{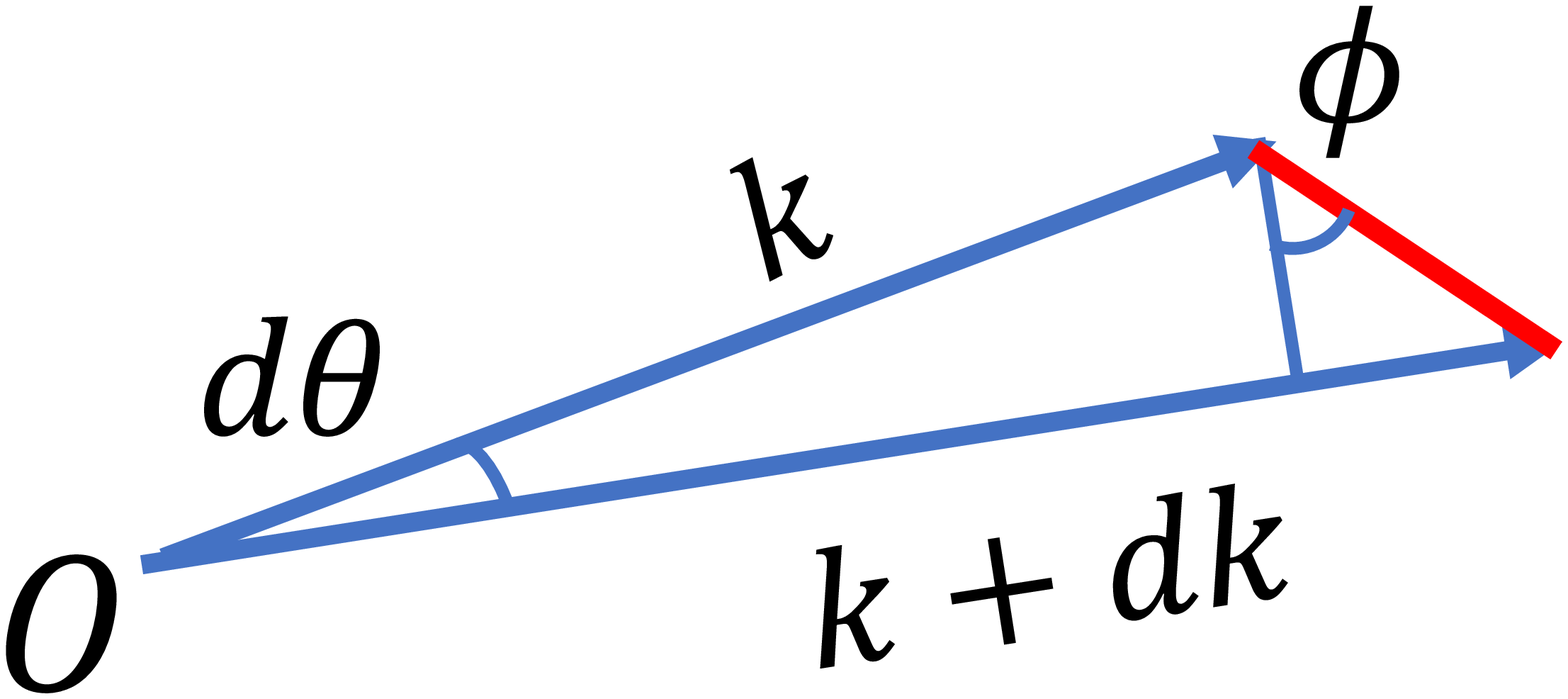}\label{fig:theta}}
  \caption{(a) For a closed Fermi pocket, there must be at least two points at which the momentum is perpendicular to the Fermi surface, and hence is parallel with the velocity. (b) The function of the Fermi surface $k(\theta)$ has at least two extrema for a closed Fermi pocket. (c) $\bv_t$ is rotated from $\bv_n$ by $\phi$, and hence the segment of the Fermi surface (red) is rotated from the PBCES (green dashed) by $\phi$. (d) If the PBCESs (green dashed) are elliptical, $k(\theta)$ could have a local maximum $k_{\text{lm}}$, where the Fermi surface (red) is tangential to a circle (black dotted). (e) The function $k(\theta)$ is monotonic for an isotropic Weyl point ($k_c$) and can have local extrema for a generic Weyl point ($k_g$). (f) The geometric relation to derive $\theta(\bk)$ from $\phi(\bk)$, where the red segment represents an infinitesimal part of the Fermi surface.}
\end{figure}

We give a proof by contradiction to show that there is no closed Fermi pocket in the surface Brillouin zone if the physics is governed by the isotropic Weyl Hamiltonian with band bending. Assume the surface is at the $z=0$ plane. Firstly, we assume that there is a closed Fermi pocket, which can be described by a function $k(\theta)$, where $k=\sqrt{k_x^2+k_y^2}$ and $\theta$ is the angle between $\bk=(k_x,k_y)$ and the $k_x$-axis, as shown in Fig.\ref{fig:closed}. Since $k(0)=k(2\pi)$, there must be a maximum $k_{\text{max}}$ and a minimum $k_{\text{min}}$ where $k'(\theta)=0$ and the momentum is perpendicular to the Fermi surface, as shown in Fig.\ref{fig:k_theta}. Since the band velocity is defined as the gradient of the energy in the momentum space, i.e. $\bv_n=\nabla_\bk\mathcal{E}(\bk)$, it is perpendicular to the Fermi surface, which is a constant-enregy surface (CES). Therefore, at $k_{\text{max}}$ and $k_{\text{min}}$, the velocity is parallel with the momentum. Besides the global extrema, there may be an even number of local extrema where the momentum and the velocity are parallel. Therefore, the minimum number of points at a closed Fermi pocket where the velocity and momentum are parallel is two. If there are no such points, there cannot be a closed Fermi pocket. Next, we show that for isotropic Weyl fermions at the surface, their velocity and momentum cannot be parallel in the presence of band bending. Consider the isotropic Weyl Hamiltonian $\mathcal{H}(\bk)=v\bk\cdot{\bf\sigma}$, where $\sigma$'s are Pauli matrices and $v$ is the Fermi velocity. We have the dispersion $\mathcal{E}_\pm(\bk)=\pm vk$ and the corresponding Berry curvature ${\bf\Omega}_\pm(\bk)=\pm C\bk/(2k^3)$, where $C$ is the chirality. Without loss of generality, we consider the upper cone, which has $\mathcal{E}(\bk)=vk$ and ${\bf\Omega}(\bk)=C\bk/(2k^3)$. At the surface $z=0$, the bending potential is along the $z$ direction, so $\dot{\bk}=f(z)\hat{{\bf z}}$. For the surface states, $\bk=(k_x,k_y,0)$ since the motion in $z$-direction is suppressed. The band velocity $\bv_n=v\hat{\bk}$ is parallel with $\bk$, while the anomalous velocity $\bv_a=Cf(z)/(2k^2)\hat{\bk}\times\hat{\bf z}$ is perpendicular to $\bk$. Usually $f(z)$ has a definite sign\cite{Li2015}, corresponding to the force exerted to the electrons by the surface, and $\bv_a$ is along the direction $\text{sgn}{\left(Cf(z)\right)}\hat{\bk}\times\hat{\bf z}$. The total velocity $\bv_t=\bv_n+\bv_a$ is rotated clockwise (anticlockwise) by an angle $\phi$ from the direction of $\bk$ if $\text{sgn}(Cf(z))=1$ ($-1$), as shown in Fig.\ref{fig:phi}. Therefore, the velocity and momentum cannot be parallel, proving the absence of a closed Fermi pocket. Note that $\bv_t$ is perpendicular to the true Fermi surface, a segment of which is shown as the red curve in Fig.\ref{fig:phi}, which is rotated from the projected bulk CES (PBCES) by the same angle $\phi$.

The above proof works for the Weyl Hamiltonian without anisotropy or tilting. If the Weyl cone is anisotropic or (and) tilted, the velocity and momentum may be parallel, but there still cannot be closed Fermi pockets. For simplicity, we prove this for anisotropic Weyl fermions without tilting, but the proof can be generalized to the case with tilting immediately. As shown in Fig.\ref{fig:ellipse}, the PBCESs in this case are ellipses. At a PBCES, the band velocity $\bv_n$ itself is not parallel with $\bk$ except at a finite set of points where $\bk$ is along the semi-major or semi-minor axes of the ellipses. The total velocity $\bv_t$ is also rotated either clockwise or anticlockwise from $\bv_n$, depending on the sign of $Cf(z)$. There is a possibility that $\bv_t$ is parallel with $\bk$ at certain $\bk$ points. When this occurs, as shown in Fig.\ref{fig:ellipse}, a segment of the Fermi surface is tangential to a circle. At such a point, $k(\theta)$ has a local maximum $k_{\text{lm}}$. However, this cannot be a global maximum. As $\bv_t$ is rotated from $\bv_n$, the segment of the Fermi surface is rotated by the same angle from the elliptical PBCES at the same $\bk$ point. As $\theta$ increases, the segment penetrates to PBCESs with higher and higher energies, so that the Fermi surface cannot be closed, but gains a spiraling structure. A schematic plot of $k(\theta)$ in this case is shown as $k_g$ in Fig.\ref{fig:kg}, which is compared with the one for an isotropic Weyl point $k_c$ where no local extrema appear.

\begin{figure}[t]
  \centering
  \subfigure[]{\includegraphics[width=0.2\textwidth]{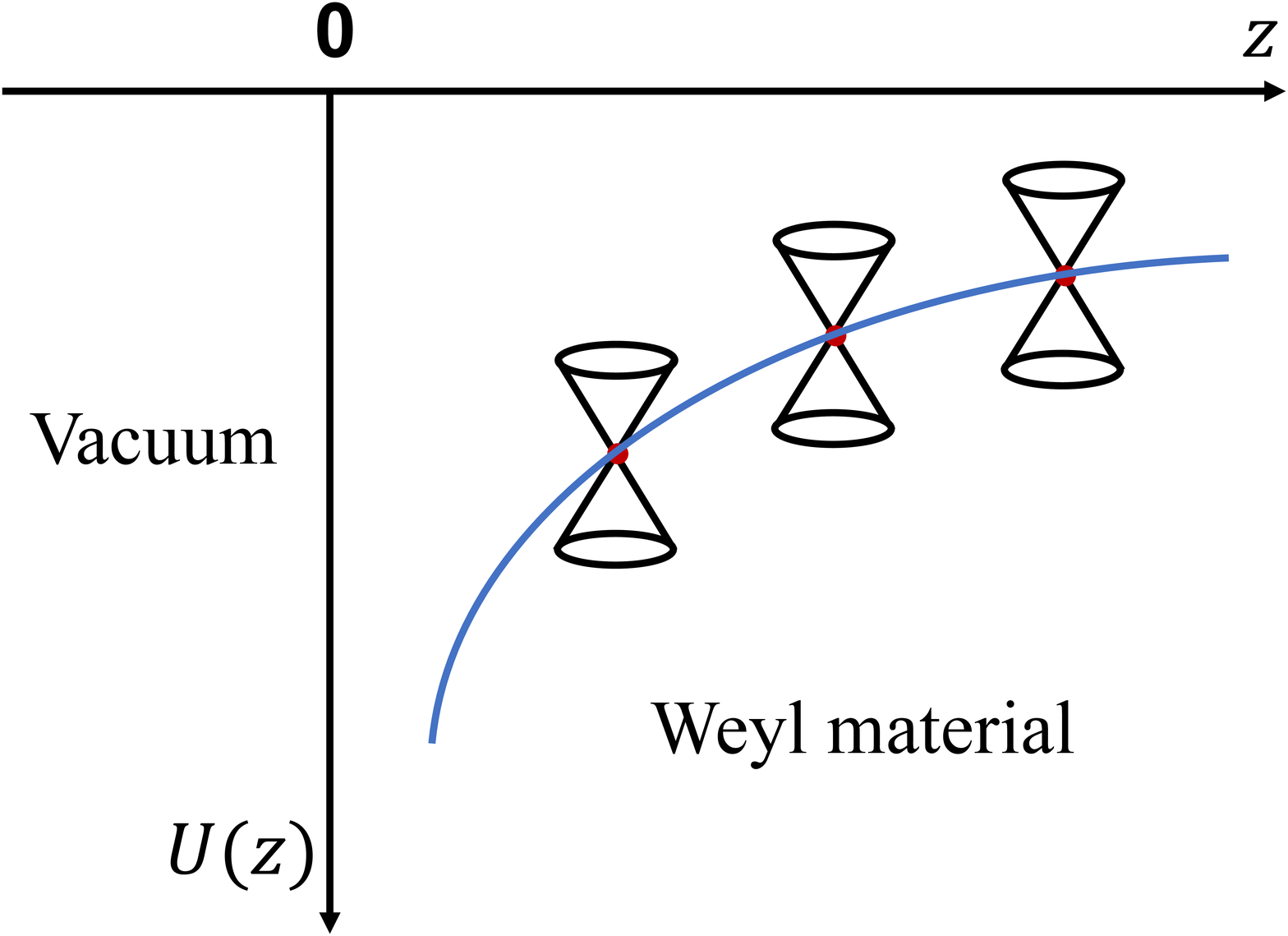}\label{fig:zpotential}}~~
  \subfigure[]{\includegraphics[width=0.2\textwidth]{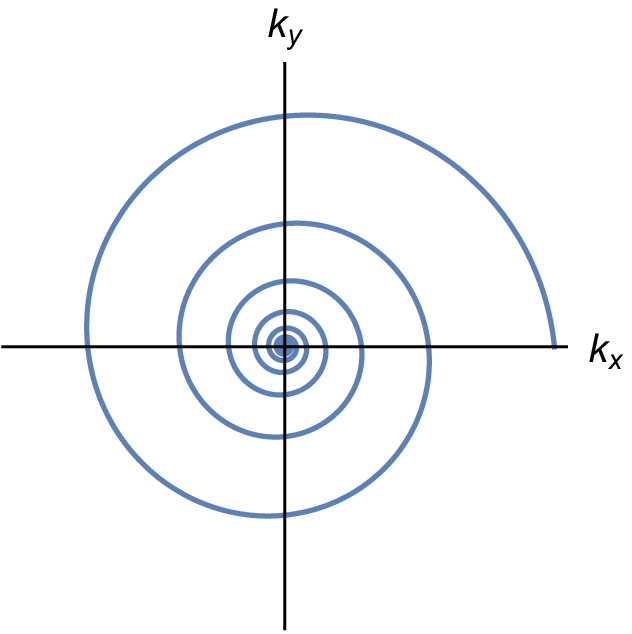}\label{fig:spiral}}
  \subfigure[]{\includegraphics[width=0.2\textwidth]{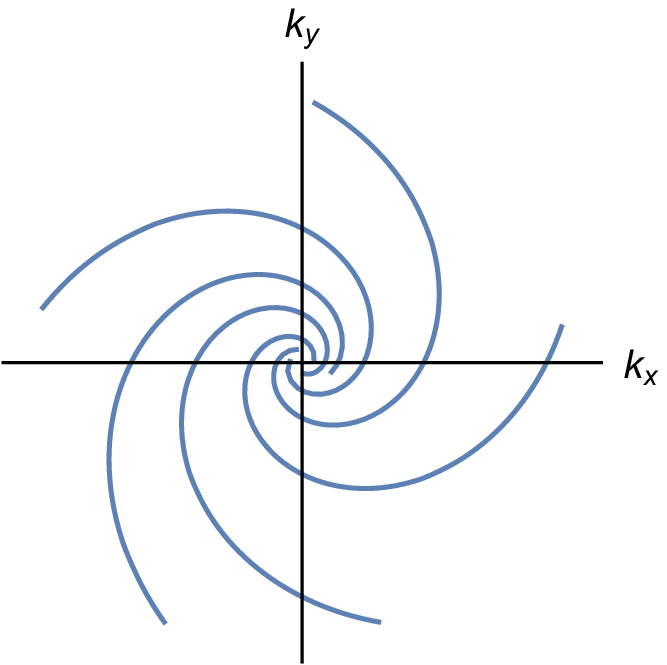}\label{fig:ces}}~~
  \subfigure[]{\includegraphics[width=0.2\textwidth]{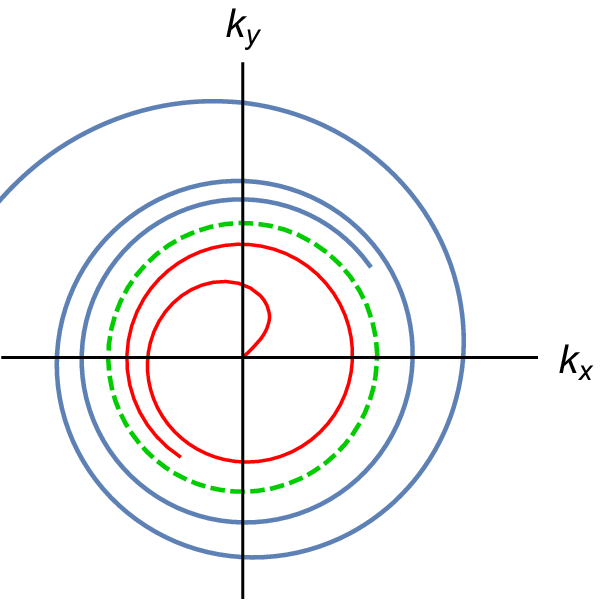}\label{fig:spurious}}
  \caption{(a) The potential due to band bending near the surface. (b) The spiraling Fermi arc of a single Weyl point at $\mu=0$. (c) CESs at $\mu=0$. (d) The spiraling Fermi arc (blue) at $\mu\neq0$, and a spurious Fermi arc (red) inside the PBFS (green dashed). The Weyl point in (b) has opposite chirality with that in (c) and (d), given $\gamma>0$.}
\end{figure}

{\it Spiraling Fermi arcs.}---Now we derive the spiraling surface Fermi arcs explicitly. We write the low energy Hamiltonian of an isotropic Weyl point with band bending in the form\cite{Li2015}
\begin{eqnarray}
H=v\mathbf{k\cdot\sigma}-\frac{\gamma}{z},
\end{eqnarray}
where $-\gamma/z$ describes the potential due to band bending, as shown in Fig.\ref{fig:zpotential}. $\gamma$ can be either positive or negative\cite{Li2015}. Instead of solving the problem quantum mechanically, we study it using the semiclassical EOMs. The energy is the sum of the kinetic part and the potential part, $\mathcal{E}(\bk,\br)=vk-\gamma/z$. The band velocity is $\bv_{n}=v\hat{\bk}$, and the anomalous velocity $\bv_{a}=-\dot{\bk}\times{\bf\Omega}=\frac{C\gamma}{2k^{2}z^{2}}\hat{\bk^{\perp}}$, where $\hat{\bk^{\perp}}=\hat{\bf z}\times\hat{\bk}$. The total velocity $\bv_t$, as well as the Fermi surface segment, is rotated anticlockwise (clockwise) by an angle $0<\phi<\pi/2$ from $\hat{\bk}$ if $\sgn(C\gamma)=1$ ($-1$). The angle $\phi$ between $\bv_t$ and $\bk$, as shown in Fig.\ref{fig:phi}, is found to be
\begin{eqnarray}
\phi(\bk)=\tan^{-1}\frac{|\mathbf{v}_a|}{|\mathbf{v}_n|}=\tan^{-1}\frac{(vk-\mu)^{2}}{2|\gamma| vk^{2}},
\end{eqnarray}
where we have fixed the energy to be $\mu$. $\phi(\bk)$ is independent of the direction of $\bk$, since the Hamiltonian is isotropic. We note two special cases: for $\mu=0$, $\phi=\tan^{-1}(v/2|\gamma|)$ is independent of $k$; and for $\mu=vk$, which corresponds to $z\to\infty$, $\phi=0$, describing the bulk states which do not have anomalous velocity.

To find the function of the Fermi surface $k(\theta)$, we use the geometric relation shown in  Fig.\ref{fig:theta}, and find
\bea
dk=-\sgn(C\gamma)kd\theta\tan\phi,
\eea
from which we get
\begin{eqnarray}
\theta=\theta_{0}-\frac{2C\gamma}{v}\left[\mu(\frac{1}{vk_{0}-\mu}-\frac{1}{vk-\mu}) +\ln\frac{vk-\mu}{vk_{0}-\mu}\right],\label{eq:theta}
\end{eqnarray}
where $k_{0}$ and $\theta_{0}$ are chosen artificially. Specifically, when $\mu=0$, we have
\begin{eqnarray}
k(\theta)=k_{0}\exp\frac{-v(\theta-\theta_{0})}{2C\gamma},
\end{eqnarray}
yielding the spiraling Fermi arc shown in Fig.\ref{fig:spiral}, which agrees with the quantum mechanical solution\cite{Li2015}. At this step one must reconcile the different symmetries of the Hamiltonian and the Fermi surface: while the Hamiltonian has rotational symmetry, the Fermi arc does not. To resolve the contradiction, we show a streamline plot of the vector field $\bv_t\times\hat{\bf z}$ in Fig.\ref{fig:ces}, which are just the CESs. Actually, there are infinite
CESs related to each other by a rotation, and they cover the whole momentum space and do not cross each other. The entirety of them restores the rotational symmetry. Any choice of $k_0>0$ and $-\infty<\theta_0<\infty$ breaks the rotational symmetry and chooses one of them to be the spiraling Fermi arc.

If the Fermi energy is at $\mu\neq0$, the spiraling surface Fermi arc winds around the projected bulk Fermi surface (PBFS) and stays outside of it, as represented by the blue curve in Fig.\ref{fig:spurious}. However, there are also solutions of Eq.\ref{eq:theta} that give rise to arcs inside of it, indicated by the red curve in Fig.\ref{fig:spurious}. Such states cannot stay at the surface, as they can tunnel into the bulk through the Klein tunneling mechanism.

We find that the semiclassical approach is applicable here no matter how large the potential is. The key point is that the smaller the gap in the momentum space is (as one approaches the projection of the Weyl point in the surface Brillouin zone), the smaller the electric field is felt due to the steeper penetration into the bulk. Therefore, this is a special problem where the semiclassical approach always applies.

\begin{figure}[t]
\centering
\subfigure[]{\includegraphics[width=0.22\textwidth]{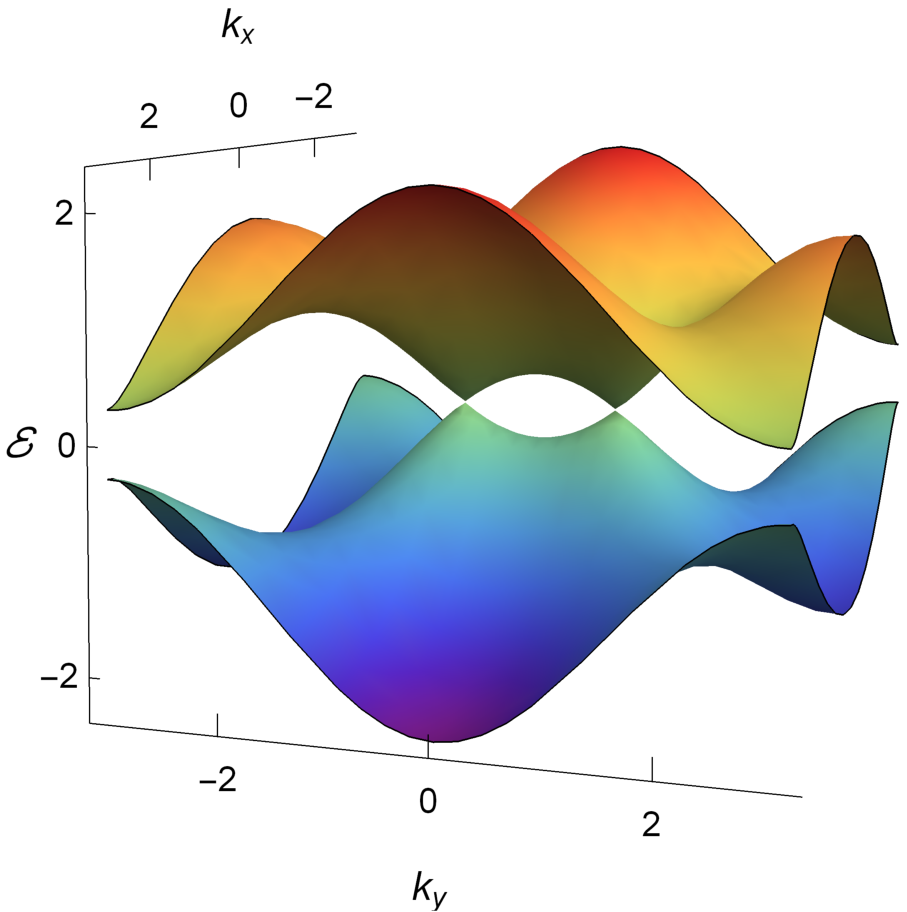}\label{fig:energy}}~~
\subfigure[]{\includegraphics[width=0.22\textwidth]{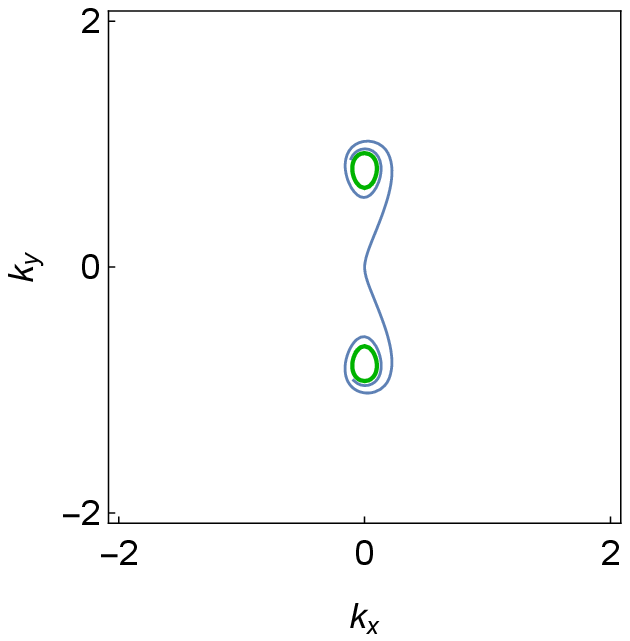}\label{fig:iso}}
\caption{(a) The dispersion with $k_z=0$. (b) The spiraling Fermi arc (blue) emanating from the PBFS (green), with the parameters  $\Delta=0.7$ and $\mu=0.1$.}\label{fig:twoWeyl}
\end{figure}

According to fermion doubling theorem\cite{Nielsen1981}, in lattice systems Weyl points with opposite chirality appear in pairs. Now we look into the case with two Weyl points. The Hamiltonian of a lattice model with two Weyl points reads $H={\bf d}\cdot{\bf \sigma}$, where
\begin{eqnarray}
{\bf d}=\left(\sin k_{x},\sin k_{z},\cos k_{y}-\Delta+(2-\cos k_{x}-\cos k_{z})\right),
\end{eqnarray}
where $\Delta$ controls the position of Weyl points. The dispersion with $k_z=0$ is shown in Fig.\ref{fig:energy}, with the two Weyl points residing at the $k_y$-axis. For the upper band, the three components of the Berry curvature are given by
\begin{eqnarray}
\Omega_{x}&=&\frac{\sin k_{x}\sin k_{y}\cos k_{z}}{2|{\bf d}|^{3}},\nonumber\\
\Omega_{y}&=&\frac{\cos k_x+\cos k_z-\cos k_{x}\cos k_{z}(2-\Delta+\cos k_{y})}{2|{\bf d}|^{3}},\nonumber\\
\Omega_{z}&=&\frac{\sin k_{z}\sin k_{y}\cos k_{x}}{2|{\bf d}|^{3}}.
\end{eqnarray}
Using the same strategy for the single Weyl point, we can derive the total velocity for the surface states subject to the bending potential $U(\br)\sim 1/z$, and hence the CESs. In Fig.\ref{fig:iso}, we show the surface Fermi arc. Near the PBFS, the spiraling structure can be recognized.

{\it Surface magnetoplasma of Dirac semimetals.}---Now we investigate a related system, a DSM in a magnetic field. Fermi arcs are also observed in DSMs\cite{Liu2014,Xiong2015,Hosur2013}. Usually the appearance of Fermi arcs is understood like this: the magnetic field couples to spin via Zeeman effect, which splits a Dirac point into two Weyl points, hence generating Fermi arc surface states. Here we take into account the orbital effect of the magnetic field by considering also the minimal coupling. Assume a confining potential at the surface $z=0$, similar to the edge of QHE, and the magnetic field ${\bf B}=B\hat{\bf y}$. Since the system with $1/z$ potential is hard to be analytically solved, we linearize the confining potential, and apply a constant electric field ${\bf E}=E\hat{\bf z}$ perpendicular to the surface. Therefore, we are treating the problem of Weyl electrons moving in crossed constant electric and magnetic fields. This problem can be solved quantum mechanically\cite{Alisultanov2017,Lukose2007}, with the Landau bands solution(See Supplemental Materials).
\bea
\e_n&=&\sgn(n)\hbar v\sqrt{\frac{2\abs{n}(1-\beta^2)^{\frac{3}{2}}}{\ell^2}+k_y^2(1-\beta^2)}+\hbar k_x v_D\nonumber\\
\eea
for $n\neq0$, and
\bea
\e_0=C\hbar v\sqrt{1-\beta^2}k_y+\hbar k_x v_D
\eea
for $n=0$, where $\ell=\sqrt{\hbar/eB}$, $v_D=E/B$ and $\beta=v_D/v$. All the Landau bands are linear in $k_x$, so that they cannot have closed Fermi surfaces. Moreover, the zeroth Landau band is also linear in $k_y$. Therefore, its Fermi surface is a line. See Supplemental Materials for details.

The fact that $\e_0$ is linear in both $k_x$ and $k_y$ reflects the chiral nature of the surface states from two origins, which can be understood intuitively from the semiclassical EOMs. It is known from special relativity that if $vB>E$, we can always boost to a frame of reference where the electric field vanishes and the magnetic field is reduced, i.e. ${\bf E}'=0$ and ${\bf B}'=B'\hat{\bf y}$ where $B'=\sqrt{1-\beta^2}B$. We can solve the semiclassical EOMs in the boosted frame, and then transform back to the original frame. In the boosted frame, a wavepacket of Weyl electrons subject to ${\bf B}'$ not only makes cyclotron motion in a plane perpendicular to the $y$-axis, but also moves along the $y$-axis even if $k_y=0$. This is because the Lorentz force combined with the Berry curvature generates the anomalous velocity along the $y$-axis. Actually, this is the origin of the chiral magnetic effect (CME)\cite{Zhou2012}. Transforming back to the original frame, the wavepacket gains a drift velocity $v_D=E/B$ along the $x$-axis, resulting in skipping orbits. Therefore, the chiral nature of the surface states originates from both the drift motion and the CME.

If the potential at the surface is very steep, most part of a skipping orbit does not feel the electric field, and at the surface the wavepacket is reflected, during which process the IF shift occurs. On average, the IF shift gives rise to a velocity in $y$-direction. Therefore, in this case, the motion of a wavepacket is composed of the drift motion, the CME and the IF shift.

{\it Summary.}---In summary, we have interpreted the morphology of the surface states of WSMs in a semiclassical way, which is more intuitive than the quantum mechanical solution. Using the semiclassical EOMs, we mapped out the spiraling Fermi arcs for a single Weyl point and a lattice with two Weyl points. We also discussed the surface Fermi arcs of Dirac semimetals in magnetic fields, and found that the drift motion, the CME and the IF shift combine to yield the Fermi arcs.

{\it Acknowledgements.}---We thank S.-K. Jian and Y.-Z. You for useful discussions. This work was supported by the National Key Research and Development Program of China (Grant Nos. 2017YFA0206203, and 2018YFA0306001), the National Natural Science Foundation of China (Grant Nos. 12004442, 11974432, and 92165204), the Guangdong Basic and Applied Basic Research Fund (Grant No. 2019A1515011337), the Shenzhen International Quantum Academy (Grant No. SIQA202102), and the Leading Talent Program of Guangdong Special Projects (Grant No. 201626003).

\bibliography{FAbib}

\begin{thebibliography}{36}%
\makeatletter
\providecommand \@ifxundefined [1]{%
 \@ifx{#1\undefined}
}%
\providecommand \@ifnum [1]{%
 \ifnum #1\expandafter \@firstoftwo
 \else \expandafter \@secondoftwo
 \fi
}%
\providecommand \@ifx [1]{%
 \ifx #1\expandafter \@firstoftwo
 \else \expandafter \@secondoftwo
 \fi
}%
\providecommand \natexlab [1]{#1}%
\providecommand \enquote  [1]{``#1''}%
\providecommand \bibnamefont  [1]{#1}%
\providecommand \bibfnamefont [1]{#1}%
\providecommand \citenamefont [1]{#1}%
\providecommand \href@noop [0]{\@secondoftwo}%
\providecommand \href [0]{\begingroup \@sanitize@url \@href}%
\providecommand \@href[1]{\@@startlink{#1}\@@href}%
\providecommand \@@href[1]{\endgroup#1\@@endlink}%
\providecommand \@sanitize@url [0]{\catcode `\\12\catcode `\$12\catcode
  `\&12\catcode `\#12\catcode `\^12\catcode `\_12\catcode `\%12\relax}%
\providecommand \@@startlink[1]{}%
\providecommand \@@endlink[0]{}%
\providecommand \url  [0]{\begingroup\@sanitize@url \@url }%
\providecommand \@url [1]{\endgroup\@href {#1}{\urlprefix }}%
\providecommand \urlprefix  [0]{URL }%
\providecommand \Eprint [0]{\href }%
\providecommand \doibase [0]{http://dx.doi.org/}%
\providecommand \selectlanguage [0]{\@gobble}%
\providecommand \bibinfo  [0]{\@secondoftwo}%
\providecommand \bibfield  [0]{\@secondoftwo}%
\providecommand \translation [1]{[#1]}%
\providecommand \BibitemOpen [0]{}%
\providecommand \bibitemStop [0]{}%
\providecommand \bibitemNoStop [0]{.\EOS\space}%
\providecommand \EOS [0]{\spacefactor3000\relax}%
\providecommand \BibitemShut  [1]{\csname bibitem#1\endcsname}%
\let\auto@bib@innerbib\@empty
\bibitem [{\citenamefont {Wan}\ \emph {et~al.}(2011)\citenamefont {Wan},
  \citenamefont {Turner}, \citenamefont {Vishwanath},\ and\ \citenamefont
  {Savrasov}}]{Wan2011}%
  \BibitemOpen
  \bibfield  {author} {\bibinfo {author} {\bibfnamefont {X.}~\bibnamefont
  {Wan}}, \bibinfo {author} {\bibfnamefont {A.~M.}\ \bibnamefont {Turner}},
  \bibinfo {author} {\bibfnamefont {A.}~\bibnamefont {Vishwanath}}, \ and\
  \bibinfo {author} {\bibfnamefont {S.~Y.}\ \bibnamefont {Savrasov}},\ }\href
  {\doibase 10.1103/PhysRevB.83.205101} {\bibfield  {journal} {\bibinfo
  {journal} {Phys. Rev. B}\ }\textbf {\bibinfo {volume} {83}},\ \bibinfo
  {pages} {205101} (\bibinfo {year} {2011})}\BibitemShut {NoStop}%
\bibitem [{\citenamefont {Witten}(2016)}]{Witten2016}%
  \BibitemOpen
  \bibfield  {author} {\bibinfo {author} {\bibfnamefont {E.}~\bibnamefont
  {Witten}},\ }\href {\doibase 10.1393/ncr/i2016-10125-3} {\bibfield  {journal}
  {\bibinfo  {journal} {Riv. Nuovo Cim.}\ }\textbf {\bibinfo {volume} {39}},\
  \bibinfo {pages} {313} (\bibinfo {year} {2016})}\BibitemShut {NoStop}%
\bibitem [{\citenamefont {Devizorova}\ and\ \citenamefont
  {Volkov}(2017)}]{Devizorova2017}%
  \BibitemOpen
  \bibfield  {author} {\bibinfo {author} {\bibfnamefont {Z.~A.}\ \bibnamefont
  {Devizorova}}\ and\ \bibinfo {author} {\bibfnamefont {V.~A.}\ \bibnamefont
  {Volkov}},\ }\href {\doibase 10.1103/PhysRevB.95.081302} {\bibfield
  {journal} {\bibinfo  {journal} {Phys. Rev. B}\ }\textbf {\bibinfo {volume}
  {95}},\ \bibinfo {pages} {081302} (\bibinfo {year} {2017})}\BibitemShut
  {NoStop}%
\bibitem [{\citenamefont {Burkov}\ and\ \citenamefont
  {Balents}(2011)}]{Burkov2011}%
  \BibitemOpen
  \bibfield  {author} {\bibinfo {author} {\bibfnamefont {A.~A.}\ \bibnamefont
  {Burkov}}\ and\ \bibinfo {author} {\bibfnamefont {L.}~\bibnamefont
  {Balents}},\ }\href {\doibase 10.1103/PhysRevLett.107.127205} {\bibfield
  {journal} {\bibinfo  {journal} {Phys. Rev. Lett.}\ }\textbf {\bibinfo
  {volume} {107}},\ \bibinfo {pages} {127205} (\bibinfo {year}
  {2011})}\BibitemShut {NoStop}%
\bibitem [{\citenamefont {Ojanen}(2013)}]{Ojanen2013}%
  \BibitemOpen
  \bibfield  {author} {\bibinfo {author} {\bibfnamefont {T.}~\bibnamefont
  {Ojanen}},\ }\href {\doibase 10.1103/PhysRevB.87.245112} {\bibfield
  {journal} {\bibinfo  {journal} {Phys. Rev. B}\ }\textbf {\bibinfo {volume}
  {87}},\ \bibinfo {pages} {245112} (\bibinfo {year} {2013})}\BibitemShut
  {NoStop}%
\bibitem [{\citenamefont {Haldane}(2014)}]{Haldane2014}%
  \BibitemOpen
  \bibfield  {author} {\bibinfo {author} {\bibfnamefont {F.~D.~M.}\
  \bibnamefont {Haldane}},\ }\href@noop {} {\  (\bibinfo {year} {2014})},\
  \Eprint {http://arxiv.org/abs/1401.0529} {arXiv:1401.0529 [cond-mat.str-el]}
  \BibitemShut {NoStop}%
\bibitem [{\citenamefont {Okugawa}\ and\ \citenamefont
  {Murakami}(2014)}]{Okugawa2014}%
  \BibitemOpen
  \bibfield  {author} {\bibinfo {author} {\bibfnamefont {R.}~\bibnamefont
  {Okugawa}}\ and\ \bibinfo {author} {\bibfnamefont {S.}~\bibnamefont
  {Murakami}},\ }\href {\doibase 10.1103/PhysRevB.89.235315} {\bibfield
  {journal} {\bibinfo  {journal} {Phys. Rev. B}\ }\textbf {\bibinfo {volume}
  {89}},\ \bibinfo {pages} {235315} (\bibinfo {year} {2014})}\BibitemShut
  {NoStop}%
\bibitem [{\citenamefont {Sun}\ \emph {et~al.}(2015)\citenamefont {Sun},
  \citenamefont {Wu},\ and\ \citenamefont {Yan}}]{Sun2015}%
  \BibitemOpen
  \bibfield  {author} {\bibinfo {author} {\bibfnamefont {Y.}~\bibnamefont
  {Sun}}, \bibinfo {author} {\bibfnamefont {S.-C.}\ \bibnamefont {Wu}}, \ and\
  \bibinfo {author} {\bibfnamefont {B.}~\bibnamefont {Yan}},\ }\href {\doibase
  10.1103/PhysRevB.92.115428} {\bibfield  {journal} {\bibinfo  {journal} {Phys.
  Rev. B}\ }\textbf {\bibinfo {volume} {92}},\ \bibinfo {pages} {115428}
  (\bibinfo {year} {2015})}\BibitemShut {NoStop}%
\bibitem [{\citenamefont {Kim}\ \emph {et~al.}(2016)\citenamefont {Kim},
  \citenamefont {Lee}, \citenamefont {Kim},\ and\ \citenamefont
  {Park}}]{Kim2016}%
  \BibitemOpen
  \bibfield  {author} {\bibinfo {author} {\bibfnamefont {K.~W.}\ \bibnamefont
  {Kim}}, \bibinfo {author} {\bibfnamefont {W.-R.}\ \bibnamefont {Lee}},
  \bibinfo {author} {\bibfnamefont {Y.~B.}\ \bibnamefont {Kim}}, \ and\
  \bibinfo {author} {\bibfnamefont {K.}~\bibnamefont {Park}},\ }\href {\doibase
  10.1038/ncomms13489} {\bibfield  {journal} {\bibinfo  {journal} {Nat.
  Commun.}\ }\textbf {\bibinfo {volume} {7}},\ \bibinfo {pages} {13489}
  (\bibinfo {year} {2016})}\BibitemShut {NoStop}%
\bibitem [{\citenamefont {Xu}\ \emph {et~al.}(2015{\natexlab{a}})\citenamefont
  {Xu}, \citenamefont {Belopolski}, \citenamefont {Alidoust}, \citenamefont
  {Neupane}, \citenamefont {Bian}, \citenamefont {Zhang}, \citenamefont
  {Sankar}, \citenamefont {Chang}, \citenamefont {Yuan}, \citenamefont {Lee},
  \citenamefont {Huang}, \citenamefont {Zheng}, \citenamefont {Ma},
  \citenamefont {Sanchez}, \citenamefont {Wang}, \citenamefont {Bansil},
  \citenamefont {Chou}, \citenamefont {Shibayev}, \citenamefont {Lin},
  \citenamefont {Jia},\ and\ \citenamefont {Hasan}}]{Xu2015Science}%
  \BibitemOpen
  \bibfield  {author} {\bibinfo {author} {\bibfnamefont {S.-Y.}\ \bibnamefont
  {Xu}}, \bibinfo {author} {\bibfnamefont {I.}~\bibnamefont {Belopolski}},
  \bibinfo {author} {\bibfnamefont {N.}~\bibnamefont {Alidoust}}, \bibinfo
  {author} {\bibfnamefont {M.}~\bibnamefont {Neupane}}, \bibinfo {author}
  {\bibfnamefont {G.}~\bibnamefont {Bian}}, \bibinfo {author} {\bibfnamefont
  {C.}~\bibnamefont {Zhang}}, \bibinfo {author} {\bibfnamefont
  {R.}~\bibnamefont {Sankar}}, \bibinfo {author} {\bibfnamefont
  {G.}~\bibnamefont {Chang}}, \bibinfo {author} {\bibfnamefont
  {Z.}~\bibnamefont {Yuan}}, \bibinfo {author} {\bibfnamefont {C.-C.}\
  \bibnamefont {Lee}}, \bibinfo {author} {\bibfnamefont {S.-M.}\ \bibnamefont
  {Huang}}, \bibinfo {author} {\bibfnamefont {H.}~\bibnamefont {Zheng}},
  \bibinfo {author} {\bibfnamefont {J.}~\bibnamefont {Ma}}, \bibinfo {author}
  {\bibfnamefont {D.~S.}\ \bibnamefont {Sanchez}}, \bibinfo {author}
  {\bibfnamefont {B.}~\bibnamefont {Wang}}, \bibinfo {author} {\bibfnamefont
  {A.}~\bibnamefont {Bansil}}, \bibinfo {author} {\bibfnamefont
  {F.}~\bibnamefont {Chou}}, \bibinfo {author} {\bibfnamefont {P.~P.}\
  \bibnamefont {Shibayev}}, \bibinfo {author} {\bibfnamefont {H.}~\bibnamefont
  {Lin}}, \bibinfo {author} {\bibfnamefont {S.}~\bibnamefont {Jia}}, \ and\
  \bibinfo {author} {\bibfnamefont {M.~Z.}\ \bibnamefont {Hasan}},\ }\href
  {\doibase 10.1126/science.aaa9297} {\bibfield  {journal} {\bibinfo  {journal}
  {Science}\ }\textbf {\bibinfo {volume} {349}},\ \bibinfo {pages} {613}
  (\bibinfo {year} {2015}{\natexlab{a}})}\BibitemShut {NoStop}%
\bibitem [{\citenamefont {Lv}\ \emph {et~al.}(2015)\citenamefont {Lv},
  \citenamefont {Weng}, \citenamefont {Fu}, \citenamefont {Wang}, \citenamefont
  {Miao}, \citenamefont {Ma}, \citenamefont {Richard}, \citenamefont {Huang},
  \citenamefont {Zhao}, \citenamefont {Chen}, \citenamefont {Fang},
  \citenamefont {Dai}, \citenamefont {Qian},\ and\ \citenamefont
  {Ding}}]{Lv2015PRX}%
  \BibitemOpen
  \bibfield  {author} {\bibinfo {author} {\bibfnamefont {B.~Q.}\ \bibnamefont
  {Lv}}, \bibinfo {author} {\bibfnamefont {H.~M.}\ \bibnamefont {Weng}},
  \bibinfo {author} {\bibfnamefont {B.~B.}\ \bibnamefont {Fu}}, \bibinfo
  {author} {\bibfnamefont {X.~P.}\ \bibnamefont {Wang}}, \bibinfo {author}
  {\bibfnamefont {H.}~\bibnamefont {Miao}}, \bibinfo {author} {\bibfnamefont
  {J.}~\bibnamefont {Ma}}, \bibinfo {author} {\bibfnamefont {P.}~\bibnamefont
  {Richard}}, \bibinfo {author} {\bibfnamefont {X.~C.}\ \bibnamefont {Huang}},
  \bibinfo {author} {\bibfnamefont {L.~X.}\ \bibnamefont {Zhao}}, \bibinfo
  {author} {\bibfnamefont {G.~F.}\ \bibnamefont {Chen}}, \bibinfo {author}
  {\bibfnamefont {Z.}~\bibnamefont {Fang}}, \bibinfo {author} {\bibfnamefont
  {X.}~\bibnamefont {Dai}}, \bibinfo {author} {\bibfnamefont {T.}~\bibnamefont
  {Qian}}, \ and\ \bibinfo {author} {\bibfnamefont {H.}~\bibnamefont {Ding}},\
  }\href {\doibase 10.1103/PhysRevX.5.031013} {\bibfield  {journal} {\bibinfo
  {journal} {Phys. Rev. X}\ }\textbf {\bibinfo {volume} {5}},\ \bibinfo {pages}
  {031013} (\bibinfo {year} {2015})}\BibitemShut {NoStop}%
\bibitem [{\citenamefont {Xu}\ \emph {et~al.}(2015{\natexlab{b}})\citenamefont
  {Xu}, \citenamefont {Alidoust}, \citenamefont {Belopolski}, \citenamefont
  {Yuan}, \citenamefont {Bian}, \citenamefont {Chang}, \citenamefont {Zheng},
  \citenamefont {Strocov}, \citenamefont {Sanchez}, \citenamefont {Chang},
  \citenamefont {Zhang}, \citenamefont {Mou}, \citenamefont {Wu}, \citenamefont
  {Huang}, \citenamefont {Lee}, \citenamefont {Huang}, \citenamefont {Wang},
  \citenamefont {Bansil}, \citenamefont {Jeng}, \citenamefont {Neupert},
  \citenamefont {Kaminski}, \citenamefont {Lin}, \citenamefont {Jia},\ and\
  \citenamefont {Zahid~Hasan}}]{Xu2015NP}%
  \BibitemOpen
  \bibfield  {author} {\bibinfo {author} {\bibfnamefont {S.-Y.}\ \bibnamefont
  {Xu}}, \bibinfo {author} {\bibfnamefont {N.}~\bibnamefont {Alidoust}},
  \bibinfo {author} {\bibfnamefont {I.}~\bibnamefont {Belopolski}}, \bibinfo
  {author} {\bibfnamefont {Z.}~\bibnamefont {Yuan}}, \bibinfo {author}
  {\bibfnamefont {G.}~\bibnamefont {Bian}}, \bibinfo {author} {\bibfnamefont
  {T.-R.}\ \bibnamefont {Chang}}, \bibinfo {author} {\bibfnamefont
  {H.}~\bibnamefont {Zheng}}, \bibinfo {author} {\bibfnamefont {V.~N.}\
  \bibnamefont {Strocov}}, \bibinfo {author} {\bibfnamefont {D.~S.}\
  \bibnamefont {Sanchez}}, \bibinfo {author} {\bibfnamefont {G.}~\bibnamefont
  {Chang}}, \bibinfo {author} {\bibfnamefont {C.}~\bibnamefont {Zhang}},
  \bibinfo {author} {\bibfnamefont {D.}~\bibnamefont {Mou}}, \bibinfo {author}
  {\bibfnamefont {Y.}~\bibnamefont {Wu}}, \bibinfo {author} {\bibfnamefont
  {L.}~\bibnamefont {Huang}}, \bibinfo {author} {\bibfnamefont {C.-C.}\
  \bibnamefont {Lee}}, \bibinfo {author} {\bibfnamefont {S.-M.}\ \bibnamefont
  {Huang}}, \bibinfo {author} {\bibfnamefont {B.}~\bibnamefont {Wang}},
  \bibinfo {author} {\bibfnamefont {A.}~\bibnamefont {Bansil}}, \bibinfo
  {author} {\bibfnamefont {H.-T.}\ \bibnamefont {Jeng}}, \bibinfo {author}
  {\bibfnamefont {T.}~\bibnamefont {Neupert}}, \bibinfo {author} {\bibfnamefont
  {A.}~\bibnamefont {Kaminski}}, \bibinfo {author} {\bibfnamefont
  {H.}~\bibnamefont {Lin}}, \bibinfo {author} {\bibfnamefont {S.}~\bibnamefont
  {Jia}}, \ and\ \bibinfo {author} {\bibfnamefont {M.}~\bibnamefont
  {Zahid~Hasan}},\ }\href {\doibase 10.1038/nphys3437} {\bibfield  {journal}
  {\bibinfo  {journal} {Nature Physics}\ }\textbf {\bibinfo {volume} {11}},\
  \bibinfo {pages} {748} (\bibinfo {year} {2015}{\natexlab{b}})}\BibitemShut
  {NoStop}%
\bibitem [{\citenamefont {Huang}\ \emph {et~al.}(2015)\citenamefont {Huang},
  \citenamefont {Xu}, \citenamefont {Belopolski}, \citenamefont {Lee},
  \citenamefont {Chang}, \citenamefont {Wang}, \citenamefont {Alidoust},
  \citenamefont {Bian}, \citenamefont {Neupane}, \citenamefont {Zhang},
  \citenamefont {Jia}, \citenamefont {Bansil}, \citenamefont {Lin},\ and\
  \citenamefont {Hasan}}]{Huang2015}%
  \BibitemOpen
  \bibfield  {author} {\bibinfo {author} {\bibfnamefont {S.-M.}\ \bibnamefont
  {Huang}}, \bibinfo {author} {\bibfnamefont {S.-Y.}\ \bibnamefont {Xu}},
  \bibinfo {author} {\bibfnamefont {I.}~\bibnamefont {Belopolski}}, \bibinfo
  {author} {\bibfnamefont {C.-C.}\ \bibnamefont {Lee}}, \bibinfo {author}
  {\bibfnamefont {G.}~\bibnamefont {Chang}}, \bibinfo {author} {\bibfnamefont
  {B.}~\bibnamefont {Wang}}, \bibinfo {author} {\bibfnamefont {N.}~\bibnamefont
  {Alidoust}}, \bibinfo {author} {\bibfnamefont {G.}~\bibnamefont {Bian}},
  \bibinfo {author} {\bibfnamefont {M.}~\bibnamefont {Neupane}}, \bibinfo
  {author} {\bibfnamefont {C.}~\bibnamefont {Zhang}}, \bibinfo {author}
  {\bibfnamefont {S.}~\bibnamefont {Jia}}, \bibinfo {author} {\bibfnamefont
  {A.}~\bibnamefont {Bansil}}, \bibinfo {author} {\bibfnamefont
  {H.}~\bibnamefont {Lin}}, \ and\ \bibinfo {author} {\bibfnamefont {M.~Z.}\
  \bibnamefont {Hasan}},\ }\href {\doibase 10.1038/ncomms8373} {\bibfield
  {journal} {\bibinfo  {journal} {Nat. Commun.}\ }\textbf {\bibinfo {volume}
  {6}},\ \bibinfo {pages} {7373} (\bibinfo {year} {2015})}\BibitemShut
  {NoStop}%
\bibitem [{\citenamefont {Yang}\ \emph
  {et~al.}(2015{\natexlab{a}})\citenamefont {Yang}, \citenamefont {Liu},
  \citenamefont {Sun}, \citenamefont {Peng}, \citenamefont {Yang},
  \citenamefont {Zhang}, \citenamefont {Zhou}, \citenamefont {Zhang},
  \citenamefont {Guo}, \citenamefont {Rahn}, \citenamefont {Prabhakaran},
  \citenamefont {Hussain}, \citenamefont {Mo}, \citenamefont {Felser},
  \citenamefont {Yan},\ and\ \citenamefont {Chen}}]{Yang2015}%
  \BibitemOpen
  \bibfield  {author} {\bibinfo {author} {\bibfnamefont {L.~X.}\ \bibnamefont
  {Yang}}, \bibinfo {author} {\bibfnamefont {Z.~K.}\ \bibnamefont {Liu}},
  \bibinfo {author} {\bibfnamefont {Y.}~\bibnamefont {Sun}}, \bibinfo {author}
  {\bibfnamefont {H.}~\bibnamefont {Peng}}, \bibinfo {author} {\bibfnamefont
  {H.~F.}\ \bibnamefont {Yang}}, \bibinfo {author} {\bibfnamefont
  {T.}~\bibnamefont {Zhang}}, \bibinfo {author} {\bibfnamefont
  {B.}~\bibnamefont {Zhou}}, \bibinfo {author} {\bibfnamefont {Y.}~\bibnamefont
  {Zhang}}, \bibinfo {author} {\bibfnamefont {Y.~F.}\ \bibnamefont {Guo}},
  \bibinfo {author} {\bibfnamefont {M.}~\bibnamefont {Rahn}}, \bibinfo {author}
  {\bibfnamefont {D.}~\bibnamefont {Prabhakaran}}, \bibinfo {author}
  {\bibfnamefont {Z.}~\bibnamefont {Hussain}}, \bibinfo {author} {\bibfnamefont
  {S.-K.}\ \bibnamefont {Mo}}, \bibinfo {author} {\bibfnamefont
  {C.}~\bibnamefont {Felser}}, \bibinfo {author} {\bibfnamefont
  {B.}~\bibnamefont {Yan}}, \ and\ \bibinfo {author} {\bibfnamefont {Y.~L.}\
  \bibnamefont {Chen}},\ }\href {\doibase 10.1038/nphys3425} {\bibfield
  {journal} {\bibinfo  {journal} {Nat. Phys.}\ }\textbf {\bibinfo {volume}
  {11}},\ \bibinfo {pages} {728} (\bibinfo {year}
  {2015}{\natexlab{a}})}\BibitemShut {NoStop}%
\bibitem [{\citenamefont {Deng}\ \emph {et~al.}(2016)\citenamefont {Deng},
  \citenamefont {Wan}, \citenamefont {Deng}, \citenamefont {Zhang},
  \citenamefont {Ding}, \citenamefont {Wang}, \citenamefont {Yan},
  \citenamefont {Huang}, \citenamefont {Zhang}, \citenamefont {Xu},
  \citenamefont {Denlinger}, \citenamefont {Fedorov}, \citenamefont {Yang},
  \citenamefont {Duan}, \citenamefont {Yao}, \citenamefont {Wu}, \citenamefont
  {Fan}, \citenamefont {Zhang}, \citenamefont {Chen},\ and\ \citenamefont
  {Zhou}}]{Deng2016NP}%
  \BibitemOpen
  \bibfield  {author} {\bibinfo {author} {\bibfnamefont {K.}~\bibnamefont
  {Deng}}, \bibinfo {author} {\bibfnamefont {G.}~\bibnamefont {Wan}}, \bibinfo
  {author} {\bibfnamefont {P.}~\bibnamefont {Deng}}, \bibinfo {author}
  {\bibfnamefont {K.}~\bibnamefont {Zhang}}, \bibinfo {author} {\bibfnamefont
  {S.}~\bibnamefont {Ding}}, \bibinfo {author} {\bibfnamefont {E.}~\bibnamefont
  {Wang}}, \bibinfo {author} {\bibfnamefont {M.}~\bibnamefont {Yan}}, \bibinfo
  {author} {\bibfnamefont {H.}~\bibnamefont {Huang}}, \bibinfo {author}
  {\bibfnamefont {H.}~\bibnamefont {Zhang}}, \bibinfo {author} {\bibfnamefont
  {Z.}~\bibnamefont {Xu}}, \bibinfo {author} {\bibfnamefont {J.}~\bibnamefont
  {Denlinger}}, \bibinfo {author} {\bibfnamefont {A.}~\bibnamefont {Fedorov}},
  \bibinfo {author} {\bibfnamefont {H.}~\bibnamefont {Yang}}, \bibinfo {author}
  {\bibfnamefont {W.}~\bibnamefont {Duan}}, \bibinfo {author} {\bibfnamefont
  {H.}~\bibnamefont {Yao}}, \bibinfo {author} {\bibfnamefont {Y.}~\bibnamefont
  {Wu}}, \bibinfo {author} {\bibfnamefont {S.}~\bibnamefont {Fan}}, \bibinfo
  {author} {\bibfnamefont {H.}~\bibnamefont {Zhang}}, \bibinfo {author}
  {\bibfnamefont {X.}~\bibnamefont {Chen}}, \ and\ \bibinfo {author}
  {\bibfnamefont {S.}~\bibnamefont {Zhou}},\ }\href {\doibase
  10.1038/nphys3871} {\bibfield  {journal} {\bibinfo  {journal} {Nature
  Physics}\ }\textbf {\bibinfo {volume} {12}},\ \bibinfo {pages} {1105}
  (\bibinfo {year} {2016})}\BibitemShut {NoStop}%
\bibitem [{\citenamefont {Lu}\ \emph {et~al.}(2015)\citenamefont {Lu},
  \citenamefont {Wang}, \citenamefont {Ye}, \citenamefont {Ran}, \citenamefont
  {Fu}, \citenamefont {Joannopoulos},\ and\ \citenamefont {Solja{\v
  c}i{\'c}}}]{Lu2015}%
  \BibitemOpen
  \bibfield  {author} {\bibinfo {author} {\bibfnamefont {L.}~\bibnamefont
  {Lu}}, \bibinfo {author} {\bibfnamefont {Z.}~\bibnamefont {Wang}}, \bibinfo
  {author} {\bibfnamefont {D.}~\bibnamefont {Ye}}, \bibinfo {author}
  {\bibfnamefont {L.}~\bibnamefont {Ran}}, \bibinfo {author} {\bibfnamefont
  {L.}~\bibnamefont {Fu}}, \bibinfo {author} {\bibfnamefont {J.~D.}\
  \bibnamefont {Joannopoulos}}, \ and\ \bibinfo {author} {\bibfnamefont
  {M.}~\bibnamefont {Solja{\v c}i{\'c}}},\ }\href {\doibase
  10.1126/science.aaa9273} {\bibfield  {journal} {\bibinfo  {journal}
  {Science}\ }\textbf {\bibinfo {volume} {349}},\ \bibinfo {pages} {622}
  (\bibinfo {year} {2015})}\BibitemShut {NoStop}%
\bibitem [{\citenamefont {Cheng}\ \emph {et~al.}(2020)\citenamefont {Cheng},
  \citenamefont {Gao}, \citenamefont {Bi}, \citenamefont {Liu}, \citenamefont
  {Li}, \citenamefont {Guo}, \citenamefont {Yang}, \citenamefont {You},
  \citenamefont {Feng}, \citenamefont {Sun}, \citenamefont {Tian},
  \citenamefont {Chen},\ and\ \citenamefont {Zhang}}]{Cheng2020}%
  \BibitemOpen
  \bibfield  {author} {\bibinfo {author} {\bibfnamefont {H.}~\bibnamefont
  {Cheng}}, \bibinfo {author} {\bibfnamefont {W.}~\bibnamefont {Gao}}, \bibinfo
  {author} {\bibfnamefont {Y.}~\bibnamefont {Bi}}, \bibinfo {author}
  {\bibfnamefont {W.}~\bibnamefont {Liu}}, \bibinfo {author} {\bibfnamefont
  {Z.}~\bibnamefont {Li}}, \bibinfo {author} {\bibfnamefont {Q.}~\bibnamefont
  {Guo}}, \bibinfo {author} {\bibfnamefont {Y.}~\bibnamefont {Yang}}, \bibinfo
  {author} {\bibfnamefont {O.}~\bibnamefont {You}}, \bibinfo {author}
  {\bibfnamefont {J.}~\bibnamefont {Feng}}, \bibinfo {author} {\bibfnamefont
  {H.}~\bibnamefont {Sun}}, \bibinfo {author} {\bibfnamefont {J.}~\bibnamefont
  {Tian}}, \bibinfo {author} {\bibfnamefont {S.}~\bibnamefont {Chen}}, \ and\
  \bibinfo {author} {\bibfnamefont {S.}~\bibnamefont {Zhang}},\ }\href@noop {}
  {\  (\bibinfo {year} {2020})},\ \Eprint {http://arxiv.org/abs/2008.05252}
  {arXiv:2008.05252 [cond-mat.mes-hall]} \BibitemShut {NoStop}%
\bibitem [{\citenamefont {Luo}\ \emph {et~al.}(2018)\citenamefont {Luo},
  \citenamefont {Yu},\ and\ \citenamefont {Weng}}]{Weng2018Re}%
  \BibitemOpen
  \bibfield  {author} {\bibinfo {author} {\bibfnamefont {K.}~\bibnamefont
  {Luo}}, \bibinfo {author} {\bibfnamefont {R.}~\bibnamefont {Yu}}, \ and\
  \bibinfo {author} {\bibfnamefont {H.}~\bibnamefont {Weng}},\ }\href
  {https://doi.org/10.1155/2018/6793752} {\bibfield  {journal} {\bibinfo
  {journal} {Research}\ }\textbf {\bibinfo {volume} {2018}},\ \bibinfo {pages}
  {10} (\bibinfo {year} {2018})}\BibitemShut {NoStop}%
\bibitem [{\citenamefont {Yang}\ \emph {et~al.}(2011)\citenamefont {Yang},
  \citenamefont {Lu},\ and\ \citenamefont {Ran}}]{Yang2011}%
  \BibitemOpen
  \bibfield  {author} {\bibinfo {author} {\bibfnamefont {K.-Y.}\ \bibnamefont
  {Yang}}, \bibinfo {author} {\bibfnamefont {Y.-M.}\ \bibnamefont {Lu}}, \ and\
  \bibinfo {author} {\bibfnamefont {Y.}~\bibnamefont {Ran}},\ }\href {\doibase
  10.1103/PhysRevB.84.075129} {\bibfield  {journal} {\bibinfo  {journal} {Phys.
  Rev. B}\ }\textbf {\bibinfo {volume} {84}},\ \bibinfo {pages} {075129}
  (\bibinfo {year} {2011})}\BibitemShut {NoStop}%
\bibitem [{\citenamefont {Aji}(2012)}]{Aji2012}%
  \BibitemOpen
  \bibfield  {author} {\bibinfo {author} {\bibfnamefont {V.}~\bibnamefont
  {Aji}},\ }\href {\doibase 10.1103/PhysRevB.85.241101} {\bibfield  {journal}
  {\bibinfo  {journal} {Phys. Rev. B}\ }\textbf {\bibinfo {volume} {85}},\
  \bibinfo {pages} {241101} (\bibinfo {year} {2012})}\BibitemShut {NoStop}%
\bibitem [{\citenamefont {Son}\ and\ \citenamefont {Spivak}(2013)}]{Son2013}%
  \BibitemOpen
  \bibfield  {author} {\bibinfo {author} {\bibfnamefont {D.~T.}\ \bibnamefont
  {Son}}\ and\ \bibinfo {author} {\bibfnamefont {B.~Z.}\ \bibnamefont
  {Spivak}},\ }\href {\doibase 10.1103/PhysRevB.88.104412} {\bibfield
  {journal} {\bibinfo  {journal} {Phys. Rev. B}\ }\textbf {\bibinfo {volume}
  {88}},\ \bibinfo {pages} {104412} (\bibinfo {year} {2013})}\BibitemShut
  {NoStop}%
\bibitem [{\citenamefont {Xiong}\ \emph {et~al.}(2015)\citenamefont {Xiong},
  \citenamefont {Kushwaha}, \citenamefont {Liang}, \citenamefont {Krizan},
  \citenamefont {Wang}, \citenamefont {Cava},\ and\ \citenamefont
  {Ong}}]{Xiong2015}%
  \BibitemOpen
  \bibfield  {author} {\bibinfo {author} {\bibfnamefont {J.}~\bibnamefont
  {Xiong}}, \bibinfo {author} {\bibfnamefont {S.~K.}\ \bibnamefont {Kushwaha}},
  \bibinfo {author} {\bibfnamefont {T.}~\bibnamefont {Liang}}, \bibinfo
  {author} {\bibfnamefont {J.~W.}\ \bibnamefont {Krizan}}, \bibinfo {author}
  {\bibfnamefont {W.}~\bibnamefont {Wang}}, \bibinfo {author} {\bibfnamefont
  {R.~J.}\ \bibnamefont {Cava}}, \ and\ \bibinfo {author} {\bibfnamefont
  {N.~P.}\ \bibnamefont {Ong}},\ }\href@noop {} {\  (\bibinfo {year} {2015})},\
  \Eprint {http://arxiv.org/abs/1503.08179} {arXiv:1503.08179
  [cond-mat.str-el]} \BibitemShut {NoStop}%
\bibitem [{\citenamefont {Hosur}\ and\ \citenamefont {Qi}(2013)}]{Hosur2013}%
  \BibitemOpen
  \bibfield  {author} {\bibinfo {author} {\bibfnamefont {P.}~\bibnamefont
  {Hosur}}\ and\ \bibinfo {author} {\bibfnamefont {X.}~\bibnamefont {Qi}},\
  }\href {\doibase https://doi.org/10.1016/j.crhy.2013.10.010} {\bibfield
  {journal} {\bibinfo  {journal} {Comptes Rendus Physique}\ }\textbf {\bibinfo
  {volume} {14}},\ \bibinfo {pages} {857 } (\bibinfo {year}
  {2013})}\BibitemShut {NoStop}%
\bibitem [{\citenamefont {Jiang}\ \emph {et~al.}(2015)\citenamefont {Jiang},
  \citenamefont {Jiang}, \citenamefont {Liu}, \citenamefont {Sun},\ and\
  \citenamefont {Xie}}]{Jiang2015}%
  \BibitemOpen
  \bibfield  {author} {\bibinfo {author} {\bibfnamefont {Q.-D.}\ \bibnamefont
  {Jiang}}, \bibinfo {author} {\bibfnamefont {H.}~\bibnamefont {Jiang}},
  \bibinfo {author} {\bibfnamefont {H.}~\bibnamefont {Liu}}, \bibinfo {author}
  {\bibfnamefont {Q.-F.}\ \bibnamefont {Sun}}, \ and\ \bibinfo {author}
  {\bibfnamefont {X.~C.}\ \bibnamefont {Xie}},\ }\href {\doibase
  10.1103/PhysRevLett.115.156602} {\bibfield  {journal} {\bibinfo  {journal}
  {Phys. Rev. Lett.}\ }\textbf {\bibinfo {volume} {115}},\ \bibinfo {pages}
  {156602} (\bibinfo {year} {2015})}\BibitemShut {NoStop}%
\bibitem [{\citenamefont {Yang}\ \emph
  {et~al.}(2015{\natexlab{b}})\citenamefont {Yang}, \citenamefont {Pan},\ and\
  \citenamefont {Zhang}}]{Yang2015PRL}%
  \BibitemOpen
  \bibfield  {author} {\bibinfo {author} {\bibfnamefont {S.~A.}\ \bibnamefont
  {Yang}}, \bibinfo {author} {\bibfnamefont {H.}~\bibnamefont {Pan}}, \ and\
  \bibinfo {author} {\bibfnamefont {F.}~\bibnamefont {Zhang}},\ }\href
  {\doibase 10.1103/PhysRevLett.115.156603} {\bibfield  {journal} {\bibinfo
  {journal} {Phys. Rev. Lett.}\ }\textbf {\bibinfo {volume} {115}},\ \bibinfo
  {pages} {156603} (\bibinfo {year} {2015}{\natexlab{b}})}\BibitemShut
  {NoStop}%
\bibitem [{\citenamefont {Wang}\ and\ \citenamefont {Jian}(2017)}]{Wang2017}%
  \BibitemOpen
  \bibfield  {author} {\bibinfo {author} {\bibfnamefont {L.}~\bibnamefont
  {Wang}}\ and\ \bibinfo {author} {\bibfnamefont {S.-K.}\ \bibnamefont
  {Jian}},\ }\href {\doibase 10.1103/PhysRevB.96.115448} {\bibfield  {journal}
  {\bibinfo  {journal} {Phys. Rev. B}\ }\textbf {\bibinfo {volume} {96}},\
  \bibinfo {pages} {115448} (\bibinfo {year} {2017})}\BibitemShut {NoStop}%
\bibitem [{\citenamefont {Xiao}\ \emph {et~al.}(2010)\citenamefont {Xiao},
  \citenamefont {Chang},\ and\ \citenamefont {Niu}}]{Xiao2010}%
  \BibitemOpen
  \bibfield  {author} {\bibinfo {author} {\bibfnamefont {D.}~\bibnamefont
  {Xiao}}, \bibinfo {author} {\bibfnamefont {M.-C.}\ \bibnamefont {Chang}}, \
  and\ \bibinfo {author} {\bibfnamefont {Q.}~\bibnamefont {Niu}},\ }\href
  {\doibase 10.1103/RevModPhys.82.1959} {\bibfield  {journal} {\bibinfo
  {journal} {Rev. Mod. Phys.}\ }\textbf {\bibinfo {volume} {82}},\ \bibinfo
  {pages} {1959} (\bibinfo {year} {2010})}\BibitemShut {NoStop}%
\bibitem [{\citenamefont {Sinitsyn}(2007)}]{Sinitsyn2007}%
  \BibitemOpen
  \bibfield  {author} {\bibinfo {author} {\bibfnamefont {N.~A.}\ \bibnamefont
  {Sinitsyn}},\ }\href {\doibase 10.1088/0953-8984/20/02/023201} {\bibfield
  {journal} {\bibinfo  {journal} {J. Phys. Condens. Matter}\ }\textbf {\bibinfo
  {volume} {20}},\ \bibinfo {pages} {023201} (\bibinfo {year}
  {2007})}\BibitemShut {NoStop}%
\bibitem [{\citenamefont {Shi}\ \emph {et~al.}(2013)\citenamefont {Shi},
  \citenamefont {Zhang},\ and\ \citenamefont {Chang}}]{Shi2013}%
  \BibitemOpen
  \bibfield  {author} {\bibinfo {author} {\bibfnamefont {L.-k.}\ \bibnamefont
  {Shi}}, \bibinfo {author} {\bibfnamefont {S.-c.}\ \bibnamefont {Zhang}}, \
  and\ \bibinfo {author} {\bibfnamefont {K.}~\bibnamefont {Chang}},\ }\href
  {\doibase 10.1103/PhysRevB.87.161115} {\bibfield  {journal} {\bibinfo
  {journal} {Phys. Rev. B}\ }\textbf {\bibinfo {volume} {87}},\ \bibinfo
  {pages} {161115} (\bibinfo {year} {2013})}\BibitemShut {NoStop}%
\bibitem [{\citenamefont {Ferreira}\ \emph {et~al.}(2018)\citenamefont
  {Ferreira}, \citenamefont {Maciel}, \citenamefont {Penteado},\ and\
  \citenamefont {Egues}}]{Ferreira2018}%
  \BibitemOpen
  \bibfield  {author} {\bibinfo {author} {\bibfnamefont {G.~J.}\ \bibnamefont
  {Ferreira}}, \bibinfo {author} {\bibfnamefont {R.~P.}\ \bibnamefont
  {Maciel}}, \bibinfo {author} {\bibfnamefont {P.~H.}\ \bibnamefont
  {Penteado}}, \ and\ \bibinfo {author} {\bibfnamefont {J.~C.}\ \bibnamefont
  {Egues}},\ }\href {\doibase 10.1103/PhysRevB.98.165120} {\bibfield  {journal}
  {\bibinfo  {journal} {Phys. Rev. B}\ }\textbf {\bibinfo {volume} {98}},\
  \bibinfo {pages} {165120} (\bibinfo {year} {2018})}\BibitemShut {NoStop}%
\bibitem [{\citenamefont {Li}\ and\ \citenamefont {Andreev}(2015)}]{Li2015}%
  \BibitemOpen
  \bibfield  {author} {\bibinfo {author} {\bibfnamefont {S.}~\bibnamefont
  {Li}}\ and\ \bibinfo {author} {\bibfnamefont {A.~V.}\ \bibnamefont
  {Andreev}},\ }\href {\doibase 10.1103/PhysRevB.92.201107} {\bibfield
  {journal} {\bibinfo  {journal} {Phys. Rev. B}\ }\textbf {\bibinfo {volume}
  {92}},\ \bibinfo {pages} {201107} (\bibinfo {year} {2015})}\BibitemShut
  {NoStop}%
\bibitem [{\citenamefont {Nielsen}\ and\ \citenamefont
  {Ninomiya}(1981)}]{Nielsen1981}%
  \BibitemOpen
  \bibfield  {author} {\bibinfo {author} {\bibfnamefont {H.}~\bibnamefont
  {Nielsen}}\ and\ \bibinfo {author} {\bibfnamefont {M.}~\bibnamefont
  {Ninomiya}},\ }\href {\doibase https://doi.org/10.1016/0550-3213(81)90361-8}
  {\bibfield  {journal} {\bibinfo  {journal} {Nuclear Physics B}\ }\textbf
  {\bibinfo {volume} {185}},\ \bibinfo {pages} {20 } (\bibinfo {year}
  {1981})}\BibitemShut {NoStop}%
\bibitem [{\citenamefont {Liu}\ \emph {et~al.}(2014)\citenamefont {Liu},
  \citenamefont {Zhou}, \citenamefont {Zhang}, \citenamefont {Wang},
  \citenamefont {Weng}, \citenamefont {Prabhakaran}, \citenamefont {Mo},
  \citenamefont {Shen}, \citenamefont {Fang}, \citenamefont {Dai},
  \citenamefont {Hussain},\ and\ \citenamefont {Chen}}]{Liu2014}%
  \BibitemOpen
  \bibfield  {author} {\bibinfo {author} {\bibfnamefont {Z.~K.}\ \bibnamefont
  {Liu}}, \bibinfo {author} {\bibfnamefont {B.}~\bibnamefont {Zhou}}, \bibinfo
  {author} {\bibfnamefont {Y.}~\bibnamefont {Zhang}}, \bibinfo {author}
  {\bibfnamefont {Z.~J.}\ \bibnamefont {Wang}}, \bibinfo {author}
  {\bibfnamefont {H.~M.}\ \bibnamefont {Weng}}, \bibinfo {author}
  {\bibfnamefont {D.}~\bibnamefont {Prabhakaran}}, \bibinfo {author}
  {\bibfnamefont {S.-K.}\ \bibnamefont {Mo}}, \bibinfo {author} {\bibfnamefont
  {Z.~X.}\ \bibnamefont {Shen}}, \bibinfo {author} {\bibfnamefont
  {Z.}~\bibnamefont {Fang}}, \bibinfo {author} {\bibfnamefont {X.}~\bibnamefont
  {Dai}}, \bibinfo {author} {\bibfnamefont {Z.}~\bibnamefont {Hussain}}, \ and\
  \bibinfo {author} {\bibfnamefont {Y.~L.}\ \bibnamefont {Chen}},\ }\href
  {\doibase 10.1126/science.1245085} {\bibfield  {journal} {\bibinfo  {journal}
  {Science}\ }\textbf {\bibinfo {volume} {343}},\ \bibinfo {pages} {864}
  (\bibinfo {year} {2014})}\BibitemShut {NoStop}%
\bibitem [{\citenamefont {Alisultanov}(2017)}]{Alisultanov2017}%
  \BibitemOpen
  \bibfield  {author} {\bibinfo {author} {\bibfnamefont {Z.~Z.}\ \bibnamefont
  {Alisultanov}},\ }\href {\doibase 10.1134/S0021364017070050} {\bibfield
  {journal} {\bibinfo  {journal} {JETP Letters}\ }\textbf {\bibinfo {volume}
  {105}},\ \bibinfo {pages} {442} (\bibinfo {year} {2017})}\BibitemShut
  {NoStop}%
\bibitem [{\citenamefont {Lukose}\ \emph {et~al.}(2007)\citenamefont {Lukose},
  \citenamefont {Shankar},\ and\ \citenamefont {Baskaran}}]{Lukose2007}%
  \BibitemOpen
  \bibfield  {author} {\bibinfo {author} {\bibfnamefont {V.}~\bibnamefont
  {Lukose}}, \bibinfo {author} {\bibfnamefont {R.}~\bibnamefont {Shankar}}, \
  and\ \bibinfo {author} {\bibfnamefont {G.}~\bibnamefont {Baskaran}},\ }\href
  {\doibase 10.1103/PhysRevLett.98.116802} {\bibfield  {journal} {\bibinfo
  {journal} {Phys. Rev. Lett.}\ }\textbf {\bibinfo {volume} {98}},\ \bibinfo
  {pages} {116802} (\bibinfo {year} {2007})}\BibitemShut {NoStop}%
\bibitem [{\citenamefont {Zhou}\ \emph {et~al.}(2012)\citenamefont {Zhou},
  \citenamefont {Jiang}, \citenamefont {Niu},\ and\ \citenamefont
  {Shi}}]{Zhou2012}%
  \BibitemOpen
  \bibfield  {author} {\bibinfo {author} {\bibfnamefont {J.-H.}\ \bibnamefont
  {Zhou}}, \bibinfo {author} {\bibfnamefont {H.}~\bibnamefont {Jiang}},
  \bibinfo {author} {\bibfnamefont {Q.}~\bibnamefont {Niu}}, \ and\ \bibinfo
  {author} {\bibfnamefont {J.-R.}\ \bibnamefont {Shi}},\ }\href {\doibase
  https://doi.org/10.1088/0256-307X/30/2/027101} {\bibfield  {journal}
  {\bibinfo  {journal} {Chin. Phys. Lett.}\ }\textbf {\bibinfo {volume} {30}},\
  \bibinfo {pages} {027101} (\bibinfo {year} {2012})}\BibitemShut {NoStop}%
\end{thebibliography}%
\bibliographystyle{apsrev4-1}

\end{document}


\title{Supplemental Materials for ``A semiclassical approach to surface Fermi arcs in Weyl semimetals"}

\author{Jiajia Huang}
\affiliation{State Key Laboratory of Optoelectronic Materials and Technologies, Guangdong Provincial Key Laboratory of Magnetoelectric Physics and Devices, School of Physics, Sun Yat-sen University, Guangzhou 510275, China}
\author{Luyang Wang}
\email{wangly@szu.edu.cn}
\affiliation{College of Physics and Optoelectronic Engineering, Shenzhen University, Shenzhen 518060, China}
\author{Dao-Xin Yao}
\email{yaodaox@mail.sysu.edu.cn}
\affiliation{State Key Laboratory of Optoelectronic Materials and Technologies, Guangdong Provincial Key Laboratory of Magnetoelectric Physics and Devices, School of Physics, Sun Yat-sen University, Guangzhou 510275, China}
\affiliation{International Quantum Academy, Shenzhen 518048, China}
\date{\today}
\maketitle

\renewcommand{\theequation}{S\arabic{equation}}
\setcounter{equation}{0}
\renewcommand{\thefigure}{S\arabic{figure}}
\setcounter{figure}{0}
\renewcommand{\thetable}{S\arabic{table}}
\setcounter{table}{0}



\section{I. Drift motion of two-dimensional Dirac fermions in crossed fields}\label{Sec.I}
Here we study the drift motion of two-dimensional (2D) Dirac fermions in crossed electric and magnetic fields with both quantum mechanical approach and semiclassical equations of motion. The results are compared with that of a free electron gas.
\subsection{A. Quantum mechanical approach}
Assume the 2D Dirac fermions move in the $xy$ plane, and a magnetic field $\bB$ is along the $z$-axis and an electric field $\bE$ is along the $y$-axis. The Hamiltonian is
\bea
H=v(p_x+eBy)\s_x+vp_y\s_y+eEy,
\eea
where $v$ is the velocity of Dirac fermions in zero field and $\s_i$'s are Pauli matrices, and the Landau gauge ${\bf A}=(-By,0,0)$ is used. Since $p_x$ commutes with the Hamiltonian, the eigenfunction of $H$ can be written as $\Psi(x,y)=\Phi(y)e^{ikx}$, where $k$ is the wave number of the plane wave in $x$-direction. Then $p_x$ in $H$ can be replaced by its eigenvalue $\hbar k$. Explicitly,
\bea
H&=&\left(
           \begin{array}{cc}
             eEy & v(\hbar k+eBy-ip_{y}) \\
             v(\hbar k+eBy+ip_{y}) & eEy\\
           \end{array}
         \right).
\eea
Define the magnetic length $\ell=\sqrt{\hbar/eB}$, the dimensionless variable $Y=y/\ell+k\ell$ and the dimensionless operator $k_Y=-i\partial_Y$. Let $\e_1=\hbar v/\ell$ and $\e_2=eE\ell$. Then we have
\bea
H=\e_2(Y-k\ell)+M,
\eea
where
\bea
M&=&\e_1\left(
           \begin{array}{cc}
             0 & Y-ik_{Y} \\
Y+ik_{Y} & 0\\
           \end{array}
         \right).
\eea
To solve the Schr\"odinger equation $H\Phi=\e\Phi$, we apply $M$ to both sides and get
\bea
M^{2}\Phi&=&M\left[\e-\e_2(Y-k\ell)\right]\Phi.
\eea
Since
\bea
M^2&=&\e_1^2\left(
           \begin{array}{cc}
             Y^2+k_Y^2-1 &0 \\
             0 & Y^2+k_Y^2+1\\
           \end{array}
         \right)
\eea
and $[M,Y]=-i\e_1\s_y$, we have
\bea
\left[\e_1^2(Y^2+k_Y^2)-\e_1^2\s_z\right]\Phi&=&\left[\e-\e_2(Y-k\ell)\right]^2\Phi+i\e_1\e_2\s_y\Phi.
\eea
Let $\e'=\e+\e_2 k\ell$ and $\eta=\e_2/\e_1=E/vB$. The above equation can be written as
\bea
\left[k_{Y}^{2}+\left(1-\eta^{2}\right) \left(Y+\frac{\eta}{1-\eta^{2}}\frac{\e'}{\e_1}\right)^{2}\right]\Phi
&=&\left[\frac{(\e'/\e_{1})^{2}}{1-\eta^{2}}+i\eta\s_{y}+\s_{z}\right]\Phi.
\eea
Making a similarity transformation to diagonalize the right-hand side, we have
\bea
\left[k_{Y}^{2}+\left(1-\eta^{2}\right) \left(Y+\frac{\eta}{1-\eta^{2}}\frac{\e'}{\e_1}\right)^{2}\right]U^{-1}\Phi
&=&\left[\frac{(\e'/\e_{1})^{2}}{1-\eta^{2}}+\sqrt{1-\eta^2}\s_{z}\right]U^{-1}\Phi,
\eea
in which the transformation matrix
\bea
U=\frac{1}{\sqrt{2(\sqrt{1-\eta^2}+1)}}\left(\begin{array}{cc}
\sqrt{1-\eta^{2}}+1 & -\eta \\
-\eta & \sqrt{1-\eta^{2}}+1 \\
\end{array}
\right).
\eea
Define the annihilation operator
\bea
a&=&\frac{1}{\sqrt{2}(1-\eta^{2})^{\frac{1}{4}}} \left[k_{Y}-i\sqrt{1-\eta^{2}}\left(Y+\frac{\eta}{1-\eta^{2}}\frac{\e'}{\e_1}\right)\right],
\eea
which satisfies $[a,a^\dagger]=1$. Then the equation becomes
\bea
(2a^{\dagger}a+1)U^{-1}\Phi =\left[\frac{(\e'/\e_{1})^{2}}{(1-\eta^{2})^{3/2}}+\s_z\right]U^{-1}\Phi,
\eea
from which it is straightforward to show that
\bea
\e_n'&=&\pm\e_1\sqrt{2n}(1-\eta^2)^{\frac{3}{4}},\\
\Phi_n&=&U\frac{1}{\sqrt{1+\abs{\al}^2}}\left(
                                    \begin{array}{c}
                                      h_{n} \\
                                      \alpha h_{n-1}
                                    \end{array}
                                  \right)
\eea
where $n\geq 1$, and $h_n$ is the eigenfunction of $a^\dagger a$, i.e.
\bea
h_{n}&=&\frac{(-1)^{n}}{2^{n/2}\pi^{\frac{1}{4}}\sqrt{n!}}(1-\eta^{2})^{\frac{1}{8}} H_{n}((1-\eta^{2})^{\frac{1}{4}}Y_{n})\times\exp(-\frac{\sqrt{1-\eta^{2}}}{2}Y_{n}^{2})
\eea
with $H_n$ being the $n$-th Hermite polynomial and
\bea
Y_n=Y+\frac{\eta}{1-\eta^{2}}\frac{\e_n'}{\e_1}.
\eea
To determine $\al$, we bring $\Phi_n$ back to the original Schr\"odinger equation, and after some algebra, finally we find
\bea
\e_n&=&\sgn(n)\sqrt{2\abs{n}}(1-\eta^2)^{\frac{3}{4}}\e_1-\frac{\hbar k E}{B},\\
\Phi_n&=&U\frac{1}{\sqrt{2-\delta_{n,0}}}\left(
                                    \begin{array}{c}
                                      h_{\abs{n}} \\
                                      \sgn(n) h_{\abs{n}-1}
                                    \end{array}
                                  \right),
\eea
where $n$ can have both positive and negative values. The drift motion perpendicular to both fields has a velocity
\bea
v_D=\frac{1}{\hbar}\frac{\partial \e_n}{\partial k}=-\frac{E}{B},
\eea
which is the same as that of free electrons. Another way to find the drift velocity is to find the expectation value of the velocity operator defined by $\hat{v}_i=\partial H/\partial p_i$. Using the eigenfunctions $\Phi_n$, it is straightforward to show $\langle\Phi_n|\hat{v}_y|\Phi_n\rangle=0$ and $\langle\Phi_n|\hat{v}_x|\Phi_n\rangle=-E/B$.

\subsection{B. Semiclassical equations of motion}
Now we use the semiclassical equations of motion to derive the drift velocity of 2D Dirac fermions in crossed fields. In the original frame, the EOMs read
\bea
\dot{\br}&=&\nabla_\bk\mathcal{E}(\bk)=v\hat{\bk},\\
\dot{\bk}&=&-e(\bE+\dot{\br}\times\bB),
\eea
where $\hat{\bk}$ is the direction of $\bk$. We can boost the Dirac fermions to a frame in which the electric field vanishes and the magnetic field is reduced. The boosted frame has a velocity $v_D=E/B$, and we use primed variables in this frame. Then the EOMs in this frame read
\bea
\dot{\br}'&=&\nabla_{\bk'}\mathcal{E}'(\bk')=v\hat{\bk}',\label{eq:k'}\\
\dot{\bk}'&=&-e\dot{\br}'\times\bB'.
\eea
In the primed frame, $\bB'=\sqrt{1-\beta^2}\bB$ and $\bE'=0$, where $\beta=v_D/v$. In Eq.(\ref{eq:k'}), we have used the invariance of the dispersion of massless Dirac fermions under Lorentz transformation, although relativistic Doppler shift occurs as $k_x'=k_x\sqrt{(1-\beta)/(1+\beta)}$. The solution is the cyclotron motion,
\bea
k_x'+ik_y'&=&k'e^{-i\omega't'},\\
v_x'+iv_y'&=&ve^{-i\omega't'},
\eea
where $k'=\sqrt{k_x'^2+k_y'^2}$, $\omega'=evB'/k'$ and $t'$ is the time in the primed frame. Then we can find the velocity in the original frame by a Lorentz transformation,
\bea
v_x&=&\frac{v(\cos{\omega't'}+\beta)}{1+\beta\cos\omega't'},\\
v_y&=&\frac{-v\sin\omega't'}{1+\beta\cos\omega't'}.
\eea
The speed keeps the same in different frames, i.e. $\sqrt{v_x^2+v_y^2}=\sqrt{v_x'^2+v_y'^2}=v$. The average speed in $y$-direction is zero, while the drift velocity is in $x$-direction and given by
\bea
\overline{v_x}=\frac{1}{\gamma T'}\int_{0}^{\gamma T'}v_x dt,
\eea
where $T'=2\pi/\omega'$ and $\gamma=1/\sqrt{1-\beta^2}$. Since $v_x$ is a function of $t'$, to do the integration we need to express $t'$ in terms of $t$ by solving
\bea
t=\gamma(t'+\frac{\beta}{\omega'}\sin\omega't').
\eea
Then we find
\bea
\overline{v_x}=v_D,
\eea
which is the same as the quantum mechanical result.
\subsection{C. Comparison with drift motion of free electrons}
\begin{figure}[h]
  \centering
  \includegraphics[width=0.5\textwidth]{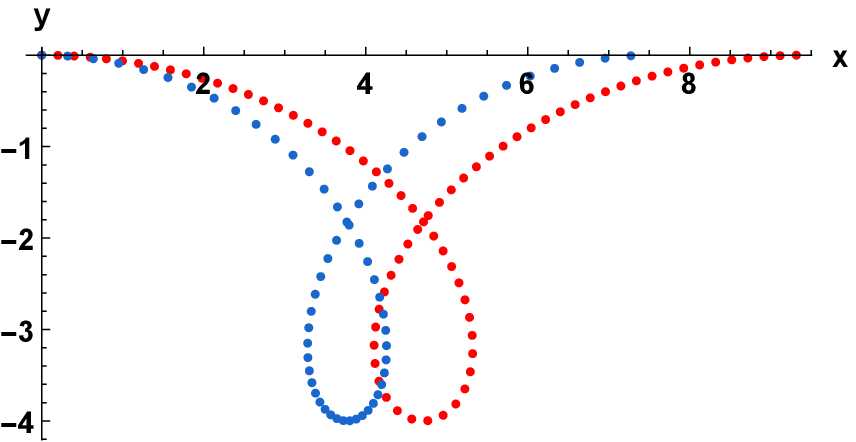}
  \caption{The drift trajectories of Dirac fermions (red dots) and free electrons (blue dots) within a period in crossed fields. The time intervals between adjacent dots are fixed.}\label{fig:trajectories}
\end{figure}

Although the drift velocity of massless Dirac fermions in crossed fields is the same as that of free electrons, their trajectories show differences. As shown in Fig.\ref{fig:trajectories}, where the red (blue) dots indicate the drift trajectory of Dirac fermions (free electrons) and are plotted with constant time intervals, the speed of Dirac fermions is constant, while the speed of free electrons varies, constrained by the conservation of energy.
\section{II. Drift motion and chiral magnetic effect of Weyl fermions in crossed fields}
\begin{figure}[h]
  \centering
  \includegraphics[width=0.8\textwidth]{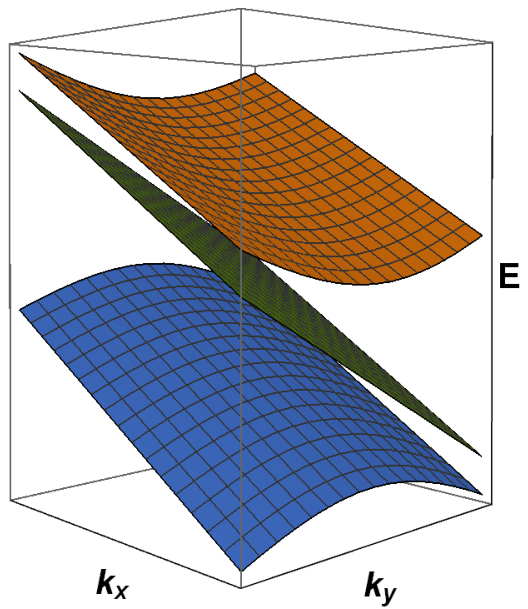}
  \caption{The Landau bands with index $n=-1,0,1$.}\label{fig:Landau}
\end{figure}

Now we study the three-dimensional (3D) Weyl fermions in crossed fields using the quantum mechanical approach. To be consistent with the main text, we assume the magnetic field $\bB$ is along the $y$-axis, while the electric field $\bE$ is along the $z$-axis. Using Landau gauge ${\bf A}=(Bz,0,0)$, we have the Hamiltonian
\bea
H_W=v(p_x-eBz)\s_x+vp_y\s_y+vp_z\s_z+eEz.
\eea
Both $p_x$ and $p_y$ commute with $H_W$, so we can replace them by their eigenvalues $\hbar k_x$ and $\hbar k_y$, respectively. It is straightforward to derive the energy eigenvalues of the Landau bands
\bea
\e_n&=&\sgn(n)\hbar v\sqrt{2\abs{n}(1-\eta^2)^{\frac{3}{2}}/\ell^2+k_y^2(1-\eta^2)}+\frac{\hbar k_x E}{B}
\eea
for $n\neq0$ and
\bea
\e_0=C\hbar v k_y\sqrt{1-\eta^2}+\frac{\hbar k_x E}{B}
\eea
for $n=0$ via the same approach used in Sec.I, where $C$ is the chirality. The drift velocity in $x$-direction is also $v_x=E/B$. The zeroth Landau band also has a constant velocity in $y$-direction, which is the origin of the chiral magnetic effect. The dispersion of Landau bands is schematically shown in Fig.\ref{fig:Landau} for three Landau bands, with $n=-1,0,1$. Obviously, the constant-energy surfaces cannot be closed.

Using a semiclassical approach, Eq. ({S20}) can be modified to have a term of anomalous velocity, which gives rise to a chiral magnetic effect. The derivation has been shown in Ref. 37 of the main text.